\documentclass[amsmath,amssymb,aps,pre,showkeys,preprint]{revtex4-1}

\usepackage{graphicx}% Include figure files
\usepackage{dcolumn}% Align table columns on decimal point
\usepackage{bm}% bold math
\usepackage{color}
\definecolor{ggreen}{rgb}{0,0.7,0.1}
\newcommand{\er}{\color{black}}

\def\bigskip{\vskip15pt}
\def\medskip{\vskip8pt}
\def\smallskip{\vskip3pt}

\def\pd#1#2{\frac{\partial #1}{\partial #2}}

\def\={\!=\!}
\def\tochka{\!\cdot\!}

\begin{document}

\preprint{APS/123-QED}

\title[]{Clustering of floating tracers in weakly divergent velocity fields}

\author{Konstantin V. Koshel}
\email{kvkoshel@poi.dvo.ru}
\affiliation{ 
V.I.Il'ichev Pacific Oceanological Institute of FEB RAS, Vladivostok, Russia
}

\author{Dmitry V. Stepanov}
\email{step-nov@poi.dvo.ru}
\affiliation{ 
V.I.Il'ichev Pacific Oceanological Institute of FEB RAS, Vladivostok, Russia
}

\author{Eugene A. Ryzhov}
\email{e.ryzhov@imperial.ac.uk}
\altaffiliation[Also at ]
{V.I.Il'ichev Pacific Oceanological Institute of FEB RAS, Vladivostok, Russia}
\affiliation{ 
Department of Mathematics, Imperial College London, London, United Kingdom
}

\author{Pavel Berloff}
\email{p.berloff@imperial.ac.uk}
\affiliation{ 
Department of Mathematics, Imperial College London, London, United Kingdom
}

\author{V. I. Klyatskin}
\email{klyatskin@yandex.ru}
\affiliation{A.M. Obukhov Atmospheric Physics Institute of RAS,\\
Moscow, Russia
}

\date{\today}

\begin{abstract}
This work focuses on buoyant tracers floating on the ocean surface and treats the geostrophic and ageostriphic
surface velocities as the 2D solenoidal (non-divergent) and potential (divergent) flow components, respectively. 
We consider a random kinematic flow model and study the process of clustering, that is, aggregation of
tracers in localized spatial patches. 
An asymptotic theory exists only for strongly divergent velocity fields and predicts \textit {complete clustering},
that is, emergence in the large-time limit of spatial singularities containing all available tracers.
To extend the theory, we consider combinations of the solenoidal and potential velocity components and
explore the corresponding regimes of the clustering process.
We have found that, even for weakly divergent flows complete clustering still persists but occurs at a significantly slower rate.
For a wide range of parameters, we have analyzed this process, as well as the other type of clustering, referred to as \textit {fragmentation clustering}, and the coarse-graining effects on clustering, and interpreted the results.
For the analyses we have considered ensembles of Lagrangian particles representing the tracer, then, introduced
and applied the methodology of \textit {statistical topography}, which paves the way for systematic studies of
 clustering in progressively more complicated flows, and for both passive and floating tracers.
\end{abstract}


\maketitle

\section{Introduction}
In turbulent oceanic and atmospheric flows, the phenomenon of clustering, that is, spatial aggregation of various tracers or material objects, manifests itself in different physical situations and on different scales \cite{Okubo_1980, McComb_1990, Jacobs_etal_2016}.
More generally, clustering processes are studied in marine ecosystems \cite{Okubo_1980}, the dynamics of clouds \cite{Kostinski_Shaw_2001}, physics of porous media \cite{Dagan_1987}, palaeontology \cite{Pater_Lissauer_2001}, and cosmology \cite{Shandarin_Zeldovich_1989}, to name a few. Clustering of floating debris and plastic litter in garbage islands \cite{Law_etal_2010, Cozar_etal_2014}, and clustering of oil spills and sargassum in patches \cite{Jacobs_etal_2016} have been actively researched problems of fundamental importance and relevant to our study.

Our focus is on clustering of \textit {floating} tracers at the \textit {ocean surface} \cite{Thomas_etal_2008, Karimova_2012, Dencausse_etal_2014, Kalda_etal_2014, Haza_etal_2016, Gaube_McGillicuddy_2017, Vali_etal_2018}.
Here, being "floating" is crucial, because, as we show, passive tracers at the ocean surface cluster differently.
Apparently, the clustering process is controlled by the multiscale and transient surface velocity, and this
motivates systematic theoretical study of clustering in random, progressively more complicated and realistic,
kinematic velocity fields, and further in dynamically constrained velocity fields.

At the ocean surface, the dominant (i.e., leading-order) geostrophic velocity can be treated as non-divergent
and purely solenoidal.
The corresponding ageostrophic velocity corrections to the leading-order fields are 3D, weakly 2D-divergent at the surface, and characterized by small vertical velocities
\cite{Reznik_Grimshaw_2001, Pallas-Sanz_Viudez_2007, Zhmur2011a, Wang_et_al_2012, McWilliams_2016, Zhang_Qiu_2018}.
At the surface, this flow correction can be treated as divergent (potential) component of the combined 2D velocity field responsible for advection and clustering of material tracers.
\cite{Maxey_1987, Boffetta_etal_1994, Klyatskin_1994b, Klyatskin_Saichev_1997, Klyatskin_2003, Cressman_Goldburg_2003, Klyatskin_2004, Monchaux_etal_2010, Bourgoin_Xu_2014, Klyatskin_2015, Klyatskin_2016, Gutierrez_Aumaitre_2016}.

Asymptotic theories of clustering are developed only for random and purely divergent kinematic flows
\cite{Klyatskin_2015, Klyatskin_2016, Klyatskin_etal_1996a, Klyatskin_etal_1996b, Saichev_Woyczynski_1996}.
Extension to weakly divergent, random kinematic flows was attempted in asymptotic sense, but the corresponding predictions have never been verified, and the underlying clustering mechanisms and other properties remain poorly understood \cite{Vali_etal_2018}.
This paper aims to close this gap.

We investigate clustering in a random kinematic velocity model and use Lagrangian-particle Monte-Carlo approach.
The paper is organized as the following.
Section $2$ formulates the problem and outlines the mathematical foundations in terms of the statistical topography methodology.
Section $3$ considers the kinematic flow model and numerical implementation of the clustering analysis.
The main results are in Section $4$, followed by conclusions in Section $5$. 

%
%%%%%%%%%%%%%%%%%%%%%%%%%%%%%%%%%%%%%%%% PROBLEM FORMULATION %%%%%%%%%%%%%%%%%%%%%%%%%%%%%%%%%%%%%%%%
%
\section{Problem formulation}

The tracer dynamics at the upper ocean surface can be formulated for both {\it passive} and {\it floating}
tracers, and these dynamical descriptions are fundamentally different, thus, resulting in profoundly different clustering properties.
First, let us introduce {\it passive-tracer concentration} $C({\bf r},t)$ and {\it floating-tracer density}
$\rho({\bf r},t)$, which are the fields of interest, both varying in space and evolving in time (here, ${\bf r}$
is the position vector and $t$ is time), subject to a velocity field ${\bf u}({\bf r},t)$.
Second, let us discuss differences between $C$ and $\rho$ and explain the terminology.
The passive tracer is just a marker of fluid (and fluid particles), and its dynamics is subject to the standard
continuity equation for material tracer; in turn, this equation can be formulated for the ocean surface.
Since our study is limited to incompressible fluids (i.e., with zero 3D velocity divergence), the concentration
$C$ is also incompressible, in the sense that it does not experience velocity divergence.
The floating tracer is not passive, in the sense that it experiences the buoyancy force that keeps it floating
on the surface, therefore, it is not a marker of fluid (and fluid particles), its dynamics is subject to the
{\it continuity equation for floating tracer}, and its density can be changed on fluid particles due to
surface velocity divergence.
The latter effect can be viewed as {\it compressibility} of the floating tracer density $\rho$, and this
justifies our choice of terminology: "density" for floating (compressible) tracers, versus "concentration"
for passive (incompressible) tracers.
Note, that both tracers experience 2D divergence of the surface velocity, but this effect is fundamentally
different from the compressibility effect.

On the ocean surface, the flow velocity can be written as ${\bf u}(x,y,0,t)=({\bf U}(x,y,t), w(x,y,0,t))$,
where ${\bf U}$ and $w$ are the horizontal and vertical velocity components, respectively.
The dynamical equations governing concentration $C$ and density $\rho$ on the ocean surface are derived in
{\textbf Appendix A}:
\begin{equation}
\biggl( {\frac{\partial }{\partial t} + {\bf{U}}({\bf{R}},t)\frac{\partial }{{\partial {\bf{R}}}}}  \biggr)
C({\bf{R}},t) = \bm{\kappa} {\Delta _{\bf{R}}}C({\bf{R}},t) \, ,
\,\,\,C({\bf{R}},0) = {C_0}({\bf{R}}) \, ;
\label{diff_adv2:1}
\end{equation}
\begin{equation}
\biggl( {\frac{\partial }{{\partial t}} + \frac{\partial }{{\partial {\bf{R}}}}{\bf{U}}({\bf{R}},t)} \biggr)
\rho ({\bf{R}},t) = \bm{\kappa} {\Delta _{\bf{R}}}\rho ({\bf{R}},t) \, ,
\,\,\,\rho ({\bf{R}},0) = {\rho _0}({\bf{R}}) \, ; 
\label{diff_adv2:2}
\end{equation}
where $C_0({\bf R})$ and $\rho_0({\bf R})$ are the initial concentration and density distributions, respectively;
${\bf R}=(x,y)$ is the 2D coordinate on the surface; $\Delta_{\bf R}$ is the 2D Laplace operator; and $\bm{\kappa}$ is
the diffusivity coefficient.
For the sake of initial simplicity, in what follows we will consider the tracer dynamics in an unbounded domain
and for $\bm{\kappa=0}$, while recognizing that both boundary and diffusive effects can be important.

Although clustering of the passive-tracer concentration is important, in this study we focus on clustering
of the floating-tracer density.
In order to distinguish between the two types of tracers and, therefore, two different types of clustering, we will refer to the passive-concentration clustering as the {\it C-clustering}, and to the floating-density clustering as the {\it D-clustering}.
Since most of the presented results are about D-clustering, we will refer to it as simply {\it clustering}, and the C-clustering terminology will be retained.

In what follows the governing equations are nondimensionalized by the length scale $L_0$, the time scale $t_0$, and the density scale $\rho^*$.
The initial area occupied by the tracer is four,
%unity, 
the initial density of the tracer is unity, and the time scale $t_0$ is the discretization time step, since the random velocity is taken to be delta-correlated in time,
and there is no correlation time scale.
The velocity scale is $U^*=L_0/t_0$, and the turbulent diffusion scale is $\kappa^*=L_0^2/t_0$.
Let us now describe and discuss the flow velocity model, which is purely kinematic thus not constrained by the fluid dynamics
for the sake of initial theoretical simplicity.

%
%%%%%%%%%%%%%%%%%%%%%%%%%%% RANDOM VELOCITY FIELDS: STATISTICAL PROPERTIES %%%%%%%%%%%%%%%%%%%%%%%%%%%
%
\subsection{Random velocity fields: statistical properties}

First, we decompose the random velocity as:
\begin{equation}
    \label{vel_sp}
{\bf U}({\bf R},t)=\gamma \, {\bf U}^p({\bf R},t) +(1-\gamma) \, {\bf U}^s({\bf R},t) \, ,
\end{equation}
where superscripts $p$ and $s$ indicate the potential (i.e., potential, purely divergent) and solenoidal
(i.e., non-divergent, purely solenoidal) velocity field components, respectively; and $0 \leq \gamma \leq 1$
is the non-dimensional parameter setting their relative contributions. 
Second, we define statistical properties of the random velocity following
\cite{Klyatskin_1994b, Klyatskin_2015, Klyatskin_Koshel_2017}:
${\bf{U}}^p({\bf{R}},t)$ and ${\bf{U}}^s({\bf{R}},t)$ are time-correlated, random Gaussian fields with spatially
homogeneous, isotropic and stationary statistics.
For such random fields the spatio-temporal velocity correlation tensors are:
\begin{equation}
    \label{corr}
B_{\alpha\beta}^j({\bf R}',\eta)
=\langle U_\alpha^j({\bf R},t) \, U_\beta^j({\bf R}+{\bf R}',t+\eta) \rangle
=\int d{\bf k} \, E_{\alpha\beta}^j({\bf k},\eta) \, e^{i{\bf kR}'} \, ,
\end{equation}
where $\bm{k}=\left( k_x, \, k_y \right)$ is the wavevector, $k=|\bm{k}|$
%, and $\eta$ is the time delay
;
$\alpha$ and $\beta$ stand for $x$ and $y$, respectively; $j$ stands for $p, \,s$; ${\bf R}'$ is a 2D spatial shift and $\eta$ is a time lag; the angular brackets denote ensemble averaging over many flow realizations.
The spectral densities are chosen to be
\begin{equation}
    \label{ek}
E_{\alpha\beta}^p({\bf k},\eta)=E^p(k,\eta) \, \frac{k_{\alpha} k_{\beta}}{k^2} \, , \hskip20pt
E_{\alpha\beta}^s({\bf k},\eta)
=E^s(k,\eta) \, \left( \delta_{\alpha\beta} -\frac{k_{\alpha} k_{\beta}}{k^2} \right) \, ,
\end{equation}
where $\delta_{\alpha\beta}$ is the Kronecker delta.
The local (in space and time) velocity correlation tensor is
\begin{equation}
    \label{sig}
B_{\alpha\beta }^j({\bf 0},0)
=\langle U_{\alpha}^j({\bf R},t) \, U_{\beta}^j({\bf R},t) \rangle
=\frac{1}{2} \, \sigma_U^2 \, \delta_{\alpha\beta} \, ,
\end{equation}
where $\sigma_U^2=B_{\alpha\alpha}({\bf 0},0)=\int d{\bf k} E(k,0)$.
Let us also introduce the {\it effective diffusivities} \cite{Klyatskin_1994b, Klyatskin_2015, Klyatskin_2016}:

\begin{equation}
\label{dsp:1}
{D^p} = \int\limits_0^\infty  d \eta \int d {\bf{k}}{\mkern 1mu} {k^2}{E^p}(k,\eta )
= \int\limits_0^\infty  d \eta \left\langle
\nabla_{\bf R} {\bf U}({\bf R}, t+\eta) \, \nabla_{\bf R} {\bf U}({\bf R}, t)
                              \right\rangle , 
\end{equation}
\begin{equation}
\label{dsp:2}
{D^s} = \int\limits_0^\infty  d \eta \int d {\bf{k}}{\mkern 1mu} {k^2}{E^s}(k,\eta )
= \frac{1}{2}\int\limits_0^\infty  d \eta \left\langle
\nabla \!\times\! {\bf U}({\bf R},t+\eta) \, \nabla \!\times\! {\bf U}({\bf R},t)
                                         \right\rangle .
\end{equation}
These diffusivities arise from the time-lag-integrated correlation tensor:
\begin{equation}
B_{kl}^{eff}\left( {\bf{r}} \right) 
=\int\limits_0^\infty  {d\tau } {B_{kl}}\left( {{\bf{r}},\tau } \right),\,\,\,\,\,B_{kl}^{eff}\left( 0 \right)
={D^0}{\delta _{kl}},\,\,\,\frac{\partial }{{\partial {r_i}}}B_{kl}^{eff}\left( 0 \right) = 0 \, ,
\end{equation}
and its second derivatives:
\begin{equation}
-8\frac{{{\partial ^2}}}{{\partial {r_i}\partial {r_j}}}B_{kl}^{eff}\left( 0 \right)
={D^s}\left( {2{\delta _{kl}}{\delta _{ij}} - {\delta _{ki}}{\delta _{lj}} - {\delta _{kj}}{\delta _{li}}} \right)
+{D^p}\left( {2{\delta _{kl}}{\delta _{ij}} + {\delta _{ki}}{\delta _{lj}} + {\delta _{kj}}{\delta _{li}}} \right).
\end{equation}
%They also appear as coefficients in the corresponding Fokker-Planck equations ***FOR WHAT?! SPECIFY*** \cite{Klyatskin_1994b, Klyatskin_2015, Klyatskin_Koshel_2017}.
In combination with the statistical topography methodology, these quantities will be used in the clustering
analyses of Section $4$.

%
%%%%%%%%%%%%%%%%%%%%%%%%%%%%%%%%%%%% STATISTICAL TOPOGRAPHY %%%%%%%%%%%%%%%%%%%%%%%%%%%%%%%%%%%%
%
\subsection{Statistical topography of random fields}

A brief description of the statistical topography methodology, applied here for a quantitative description of the clustering process, follows \cite{Klyatskin_2015}.
Let density field $\rho({\bf R},t)$  be a random in time process with a probability density function (PDF)
homogeneous in space but evolving in time.
Let us assume that, from a smooth initial density distribution, the dynamics generates a population of clusters,
which are localized regions with high tracer density.
A solution corresponding to the log-normal process, which describes the clustering, has the PDF
\begin{equation}
    \label{log_norm}
P(t;\rho)=\frac{1}{2 \rho \sqrt{\pi t/\tau}} \, \exp \left\{
                                       -\frac{\ln^2(\rho e^{t/\tau}/\rho_0)}{4 t/\tau}
                                                    \right\} \, ,
\end{equation}
where $\tau=1/D^p$ is the {\it effective diffusion time scale}.
The corresponding {\it integral distribution function} is
\begin{equation}
    \label{int_norm}
\Phi(t;\rho) \equiv \int_0^{\rho}  d\rho' \, P(t;\rho')
= \Pr \left( \frac{\ln^2(\rho e^{t/\tau}/\rho_0)}{2 \sqrt{t/\tau}} \right) \, ,
\end{equation}
where $\Pr(z)$ is the probability integral
\begin{equation}
    \label{int_prob}
\Pr(z)=\frac{1}{\sqrt{2\pi}} \, \int_{-\infty}^z dx \, \exp \left\{-\frac{x^2}{2} \right\} \, .
\end{equation}
Let us also consider the {\it indicator function}
\begin{equation}
    \label{inD^fun}
\varphi({\bf R},t; \rho)=\delta(\rho({\bf R},t)-\rho) \, ,
\end{equation}
which sifts $\rho({\bf R},t)$ at given $\rho$ via the Dirac delta-function.
The total area of the regions, where density exceeds some threshold $\bar\rho$, is referred to as the
{\it clusters area} and obtained as
\begin{equation}
    \label{sq_rnd}
S(t; \bar\rho) =\int d{\bf R} \, \theta(\rho({\bf R},t)-\bar\rho)
=\int d{\bf R} \int_{\bar\rho}^{\infty} d\rho' \varphi({\bf R},t; \rho') \, ,
\end{equation}
where $\theta(\cdot)$ is the Heaviside (step) function.
The total mass of the tracer within the clusters area is referred to as the {\it clusters mass} and obtained as
\begin{equation}
    \label{m_rnd}
M(t; \bar\rho)=\int d{\bf R} \, \rho({\bf R},t) \, \theta(\rho({\bf R},t)-\bar\rho)
=\int d{\bf R} \int_{\bar\rho}^{\infty} d\rho' \, \rho' \, \varphi({\bf R},t; \rho') \, .
\end{equation}
Ensemble averaging of (\ref{sq_rnd}) and (\ref{m_rnd}) yields:
\begin{eqnarray}
    \label{av_s}
&&\left\langle S(t; \bar\rho) \right\rangle
= \int d{\bf R} \int_{\bar\rho}^{\infty}  d\rho' \, P({\bf R},t; \rho') \, ,
\\
    \label{av_m}
&&\left\langle M(t; \bar\rho) \right\rangle
= \int d{\bf R} \int_{\bar\rho}^{\infty} d\rho' \, \rho' \, P({\bf R},t; \rho') \, ,
\end{eqnarray}
because the PDF (\ref{log_norm}) is completely local (i.e., involves no lags in time and shifts in space) and
will be referred to as one-point PDF.
The ensemble mean of the indicator function (\ref{inD^fun}) ensures that:
\begin{equation}
    \label{prob_den}
P({\bf R},t; \rho)=\langle \delta(\rho({\bf R},t)-\rho) \rangle \, ,
\end{equation}
and the expressions (\ref{av_s}) and (\ref{av_m}) become:
\begin{eqnarray}
    \label{hom_sm}
&& \langle s_{hom}(t; \bar\rho) \rangle =\langle \theta(\rho({\bf R},t)-\bar\rho) \rangle
=P \{ \rho({\bf R},t) > \bar\rho \} =\int_{\bar\rho}^{\infty} d\rho' \, P(t; \rho') \, ,
\\
&&\left\langle m_{hom}(t; \bar\rho) \right\rangle
=\frac{1}{\rho_0} \, \int_{\bar\rho}^{\infty} d\rho' \, \rho' \, P(t; \rho') \, ,
\end{eqnarray}
where $s_{hom}(t; \bar\rho)$ and $m_{hom}(t; \bar\rho)$ are, respectively, the {\it specific clusters area}
and the {\it specific clusters mass} defined by the threshold $\bar\rho$
\cite{Klyatskin_2016, Klyatskin_Koshel_2017}.
Since $\rho({\bf R},t)$ is a positive-definite field, the clustering happens with probability one, and the corresponding limits,
\begin{equation}
    \label{inf_sm}
\lim_{t \to \infty} \langle s_{hom}(t; \bar\rho) \rangle \to 0 \, , \hskip20pt
\lim_{t \to \infty} \langle m_{hom}(t; \bar\rho) \rangle \to
\frac{1}{\rho_0} \, \langle \rho(t) \rangle \, ,
\end{equation}
assert that the clusters area shrinks to zero, while the clusters mass incorporates all the available tracer.
From (\ref{log_norm}) and (\ref{hom_sm}), one can derive the specific functions for the purely divergent velocity field (i.e., for $\gamma=0$):
\begin{equation}
    \label{log_sm}
\langle s_{hom}(t,\bar\rho) \rangle
=Pr \left( \frac{\ln(\rho_0 \, e^{-D^p t}/\bar\rho)}{\sqrt{2 D^p t}} \right) \, , \hskip20pt
\langle m_{hom}(t,\bar\rho) \rangle
=Pr \left( \frac{\ln(\rho_0 \, e^{ D^p t}/\bar\rho)}{\sqrt{2 D^p t}} \right) \, ,
\end{equation}
where $\rho_0$ is the initial uniform density.
For times much larger than the diffusion time scale $1/D^p$, approximate estimates can be obtained from
(\ref{log_sm}) and (\ref{int_prob}) by using the large-argument asymptotics of the probability integral
\cite{Olver_2010}:
\begin{eqnarray}
    \label{log_app}
&&\left\langle {{s_{\hom }}(t,\bar \rho )} \right\rangle  = {\rm{P}}\{ \rho ({\bf{R}},t) > \bar \rho \}  \approx \sqrt {\frac{{{\rho _{\rm{0}}}}}{{\pi \bar \rho D^p t }}} {{\rm{e}}^{{\rm{  -  }}\frac{1}{4}D^pt}},\nonumber\\
&&\left\langle {{m_{\hom }}(t,\bar \rho )} \right\rangle /{\rho _0} \approx 1 - \sqrt {\frac{{\bar \rho }}{{\pi {\rho _0}D^p t }}} {e^{ - \frac{1}{4}D^pt}}.
\end{eqnarray}
These relations describe the {\it complete clustering} that happens exponentially with probability one, and the one-point PDF and such characteristics as clusters area and mass conveniently describe the process.
These characteristics, as well as their two-point extensions \cite{Klyatskin_2016}, are used below for analysis
of clustering in previously unexplored, weakly divergent velocity fields.

We also employ the concept of {\it typical realization curve}.
Given a random process $z(t)$, this curve is defined as the median of the integral distribution function 
%$\Phi$
(\ref{int_norm}) and found as the solution $z^*(t)$ of the equation,
\begin{equation}
    \label{tip_real}
\Phi(t; z^*(t)=\int_0^{z^*(t)} dz' \, P(t; z')=\frac{1}{2} \, ,
\end{equation}
which implies that for any time $t$: \hskip10pt $P[z(t) > z^*(t)]=P[z(t) < z^*(t)]=1/2$.
Typical realization curve of the distance between two particles \cite{Klyatskin_2015} is 
\begin{equation}
    \label{tip_eq}
l^*(t)=\exp \left\{ \frac{1}{4} \, (D^s-D^p) \, t \right\} \, .
\end{equation}
When $D^s < D^p$, the average distance between particles, as given by $l^*(t)$,  attenuates exponentially in accord with the process of complete clustering.
When $D^s > D^p$, there is no complete clustering, at least for the times such that the following constraint holds \cite{Klyatskin_2003}:
\begin{equation}
    \label{constr}
\frac{1}{4} \, D^s t \ll \ln\frac{l_{corr}}{l_0} \, ,
\end{equation}
where $l_0$ is the average initial distance between the particles, and $l_{corr}$ is the spatial correlation
radius of the velocity field.

The statistical topography methodology offers a variety of useful characteristics, such as the average length
of the isoline contour corresponding to $\rho({\bf R},t)=\bar\rho$, which is estimated as
\begin{equation}
    \label{length}
\langle L(t,\rho) \rangle =L_0 \, e^{D^s t} \, ,
\end{equation}
where $L_0$ is its initial length \cite{Klyatskin_etal_1996a, Saichev_Woyczynski_1996}.
Thus, the contour length in the purely non-divergent velocity field grows exponentially, because the tracer
patch is continuously stretched and filamented.
We refer to this process as {\it tracer fragmentation}.
In a weakly divergent velocity field, the tracer fragmentation affects and modulates the process of complete
clustering, and we address this issue.

%
%%%%%%%%%%%%%%%%%%%%%%%%%%%%%%%%%%%%% KINEMATIC VELOCITY MODEL %%%%%%%%%%%%%%%%%%%%%%%%%%%%%%%%%%%%%
%
\section{Kinematic velocity model}

In this section we introduce the kinematic model of a random velocity field with a prescribed correlation tensor, and then
study tracer clustering in the model solutions.

\subsection{Model formulation}

Following \cite{Roberts_Teubner_1995, Zirbel_Cinlar_1997, Koshel_Alexandrova_1999, Klyatskin_Koshel_2017},
we use a spectral representation of the velocity field:
\begin{eqnarray}
    \label{num_vel}
&& U_\beta ^p\left( {{\bf{R}},t} \right) = {\sigma _{\bf{U}}}\int {d{\bf{k}}\left( {a\left( {k,t} \right) + ib\left( {k,t} \right)} \right)\frac{{{k_\beta }}}{k}\exp \left( {i{\bf{kR}}} \right)} ,\nonumber\\
&& U_x^s\left( {{\bf{R}},t} \right) = {\sigma _{\bf{U}}}\int {d{\bf{k}}\left( {a\left( {k,t} \right) + ib\left( {k,t} \right)} \right)\frac{{{k_y}}}{k}\exp \left( {i{\bf{kR}}} \right)} , \\
&& U_y^s\left( {{\bf{R}},t} \right) =  - {\sigma _{\bf{U}}}\int {d{\bf{k}}\left( {a\left( {k,t} \right) + ib\left( {k,t} \right)} \right)\frac{{{k_x}}}{k}\exp \left( {i{\bf{kR}}} \right)}, \nonumber
\end{eqnarray}
where $\sigma_U$ is the standard deviation of the velocity, index $\beta$ stands for $x$ and $y$; $a({\bf k},t)$
and $b({\bf k},t)$ are Gaussian, $\delta$-correlated in time, spectral coefficients, such that:
\begin{eqnarray}
    \label{num_par}
&&\left\langle {a\left( {k,t} \right)} \right\rangle  = \left\langle {b\left( {k,t} \right)} \right\rangle  = \left\langle {a\left( {{\bf{k}},t} \right)b\left( {{\bf{k'}},t'} \right)} \right\rangle  = 0,\nonumber\\
&&\left\langle {a\left( {{\bf{k}},t} \right)a\left( {{\bf{k'}},t'} \right)} \right\rangle  = \left\langle {b\left( {{\bf{k}},t} \right)b\left( {{\bf{k'}},t'} \right)} \right\rangle  = {E}\left( k \right)\delta \left( {{\bf{k}} - {{\bf{k}}^\prime }} \right)\delta \left( {t - t'} \right).
\end{eqnarray}
The velocity field (\ref{num_vel}) corresponds to the correlation tensor (\ref{corr}); and in the physical space
it is obtained by an inverse Fourier transform with a random phase.
The spectral density is prescribed in the form
%***why is there $k-l$ asymmetry?!***
\begin{equation}
    \label{spec_den}
E(k,l)=\frac{1}{2\pi} \, \frac{l^4}{4} \, k^2 \, \exp \left\{-\frac{1}{2} \, k^2 \, l^2 \right\} \, ,
\end{equation}
where $l=l_{corr}$ is the spatial correlation radius.
The effective diffusion coefficients (\ref{dsp:1}) -- (\ref{dsp:2}) ensue from (\ref{spec_den}) and (\ref{vel_sp})
by the integration:
\begin{equation}
    \label{num_dsp}
D^s=(1-\gamma)^2 \, D^0 \, , \hskip20pt D^p=\gamma^2 \, D^0 \, , \hskip20pt D^0=\frac{\sigma_U^2}{l^2} \, t_0 \, ,
\end{equation}
where $t_0$ is the time scale.

\subsection{Numerical implementation}

Given uniformly gridded velocity fields, equations (\ref{diff_adv2:1}) and (\ref{diff_adv2:2}) are solved
numerically by a simulation of the ensemble of Lagrangian particles \cite{Klyatskin_1994b, Gat_etal_1998, Frisch_etal_1998, Koshel_Alexandrova_1999, Koshel_et_al_2013, Koshel_et_al_2015}. We do not consider the effect of molecular diffusion here  ($\bm{\kappa}  = 0$), it means that the equations  (\ref{diff_adv2:1}) and (\ref{diff_adv2:2}) are first order linear PDE. The simplest means to solve such equations is the method of characteristics \cite{Klyatskin_1994b, Koshel_et_al_2015}. 
Evolution of each Lagrangian particle is governed by
\begin{eqnarray}
    \label{char}
&&\frac{{d{\bf{R}}}}{{dt}} = {\bf{U}}\left( {{\bf{R}},t} \right),\,\,\,{\mkern 1mu} \,{\bf{R}}\left( 0 \right) = {\bm{\xi }}, \nonumber \\
&&\frac{{d\rho \left( {t;{\bm{\xi }}} \right)}}{{dt}} =  - \frac{{\partial {\bf{U}}\left( {{\bf{R}},t} \right)}}{{\partial {\bf{R}}}}\rho \left( {t;{\bm{\xi }}} \right),{\mkern 1mu} \,\,\,\,\,\rho \left( {0;{\bm{\xi }}} \right) = {\rho _0}\left( {\bm{\xi }} \right); \nonumber \\
&&\frac{{dC\left( {t;{\bm{\xi }}} \right)}}{{dt}} = 0,{\mkern 1mu} \,\,\,\,\,C\left( {0;{\bm{\xi }}} \right) = {C_0}\left( {\bm{\xi }} \right).
\end{eqnarray}
Here ${\bm{\xi }}$ is the initial position of the trajectory. These equations are integrated in time using the standard Euler-Ito scheme, and the Eulerian density and concentration fields are obtained by the relations:
\begin{eqnarray}
    \label{roc}
&&\rho \left( {{\bf{R}},t} \right) = \rho \left( t \right) = \rho \left( {t;\bm{\xi} \left( {{\bf{R}};t} \right)} \right); \nonumber \\
&&C\left( {{\bf{R}},t} \right) = C\left( t \right) = C\left( {t;\bm{\xi} \left( {{\bf{R}};t} \right)} \right).
\end{eqnarray}

We use 900 tracers in each mesh sells from the initial spot of the impurity to obtain good enough Eulerian density and concentration fields.

It is worse to mention that the effect of the molecular diffusion can be considered by adding the additional vector delta-correlated process in right part of the equations (\ref{char}), see references above.

%
%%%%%%%%%%%%%%%%%%%%%%%%%%%%%%%%%%%%%%% MODELLING RESULTS %%%%%%%%%%%%%%%%%%%%%%%%%%%%%%%%%%%%%%%
%
\section{Modelling results}

In this section we focus on modelling the nondiffusive clustering in the implemented double-periodic domain
($|x| \leq 10, \hskip10pt |y| \leq 10$) with the uniform numerical grid $2048^2$.
The initial uniform distribution of Lagrangian particles is in the sub-domain $|x| \leq 1, \hskip10pt |y| \leq 1$,
and there are $900$ particles placed initially in each grid cell.
Results for the strongly and weakly divergent velocity regimes are discussed and compared.

Let us explain how the particle distributions are obtained further below.
The random velocity field within each grid cell is treated as piecewise-constant, and this ensures that the
velocity field has discontinuities responsible for the white-noise stochastic process
\cite{Koshel_Alexandrova_1999, Klyatskin_Koshel_2017}.
Density and concentration of particles are found by averaging over the grid scale.
The averaged concentration is the number of particles in the cell divided by the cell area.
The averaged density is the number of particles in the cell divided by the area which these particles occupy.
Because each particle conserves its mass, the corresponding area experiences velocity divergence and scales as
$\rho_0/\rho(t)$.
Note, that the averaging can sometimes produce spurious concentration values, therefore, some caution with
interpretation is needed (see examples towards the end of the paper).

%
%%%%%%%%%%%%%%%%%%%%%%%% PURELY DIVERGENT AND NON-DIVERGENT FLOW REGIMES %%%%%%%%%%%%%%%%%%%%%%%%
%
\subsection{Purely divergent and non-divergent flow regimes}

\begin{figure}
\centering
\includegraphics[width=6 cm]{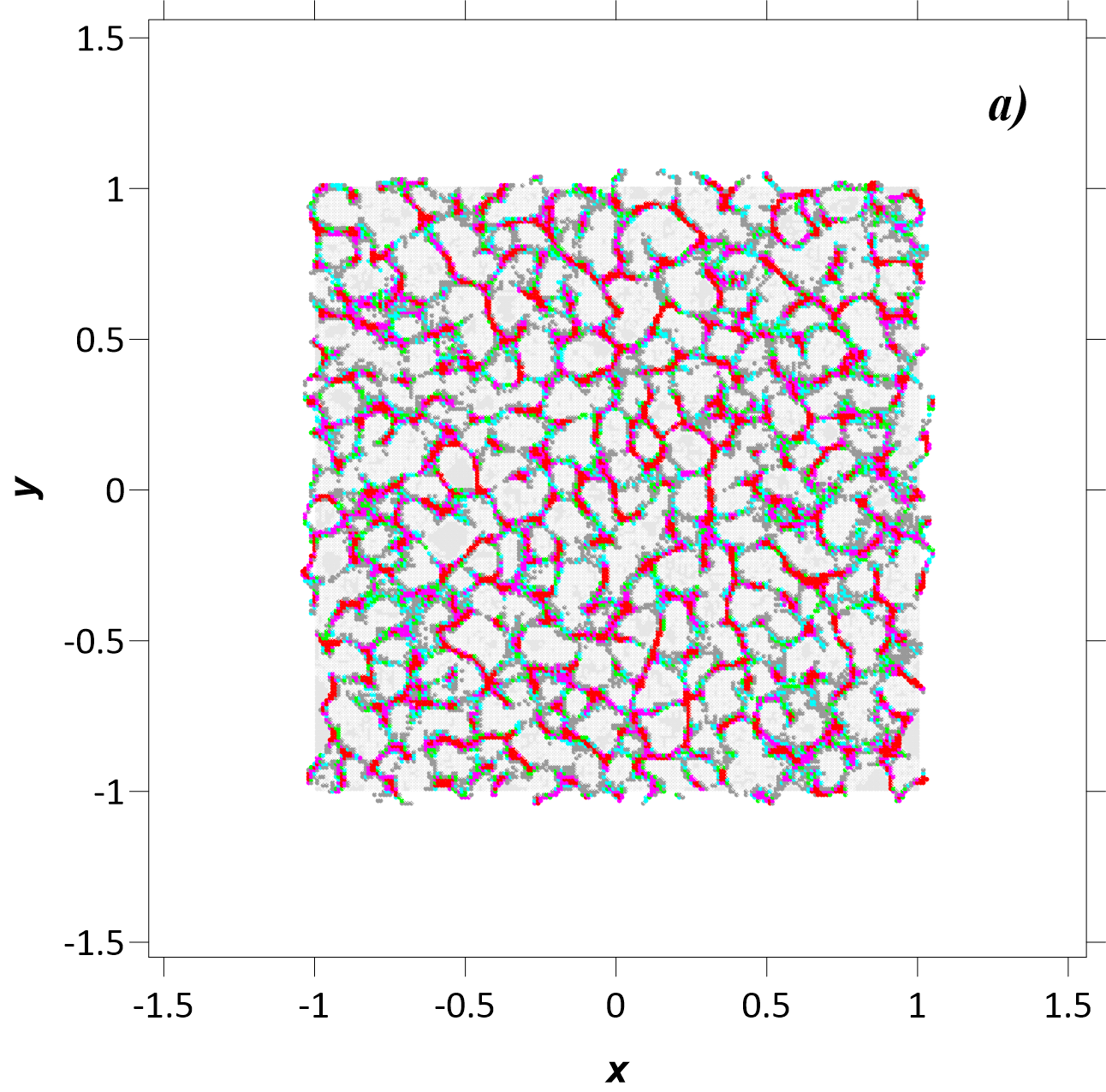}
\includegraphics[width=6 cm]{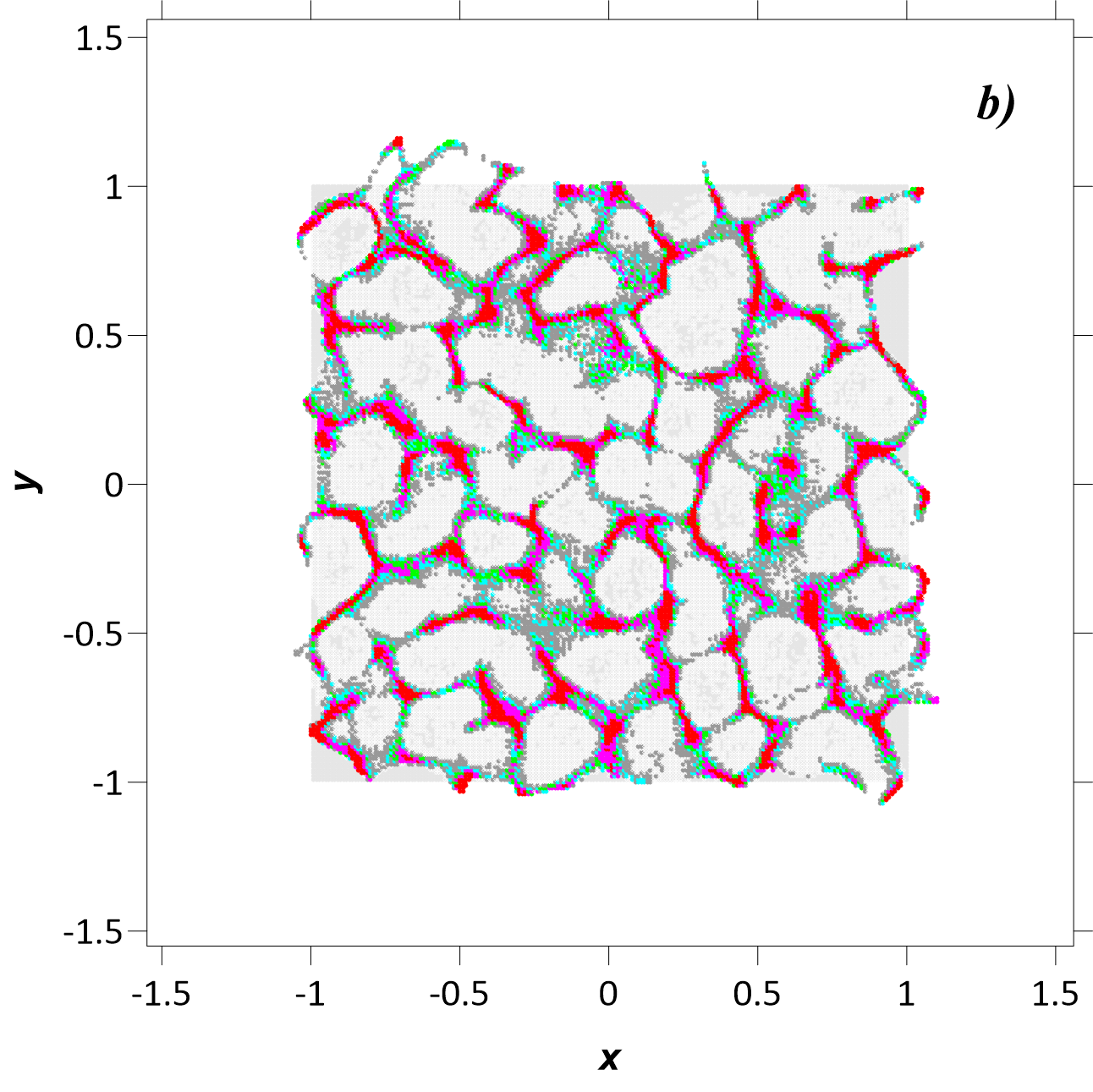}
\includegraphics[width=6 cm]{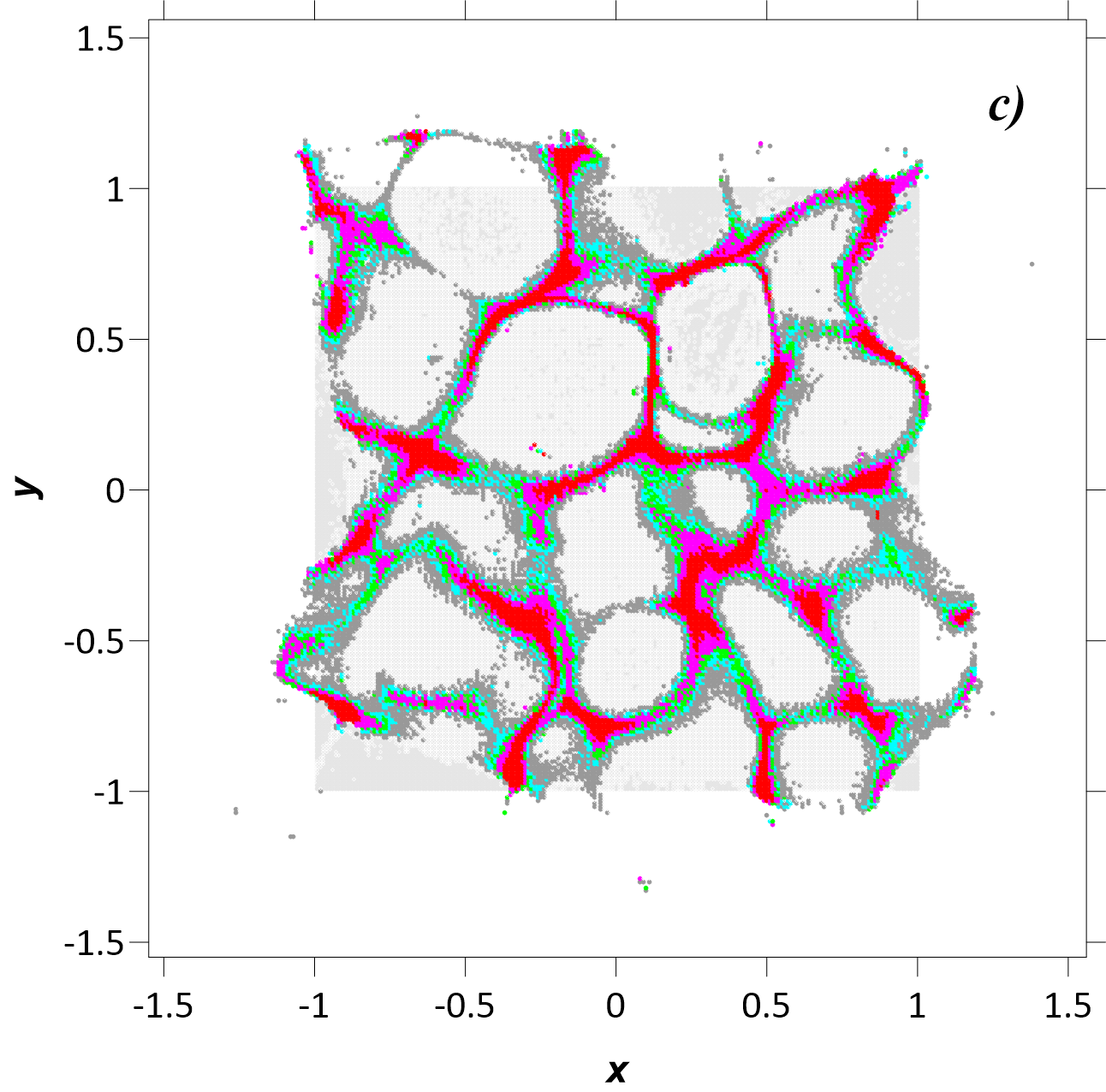}
\includegraphics[width=4 cm]{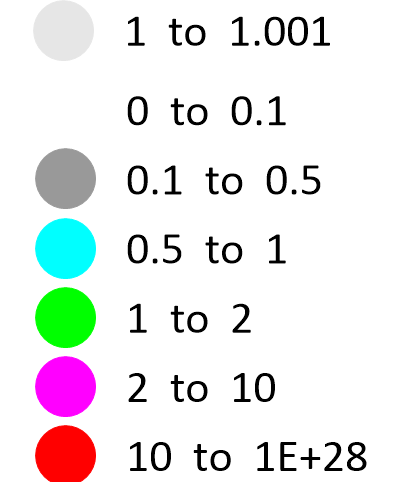}
\caption{
\label{fig1}
Density distributions for a purely divergent velocity regime ($\gamma=1$) for $t=\tau$:
(a) $l = 0.04, \hskip5pt \sigma_U=0.33$;
(b) $l = 0.08, \hskip5pt \sigma_U=0.33$;
(c) $l = 0.16, \hskip5pt \sigma_U=0.67$.
Grey square indicates the initial density distribution.
Colour scale indicates (nondimensional) values of the tracer density.
}
\vskip20pt
\end{figure}

\begin{figure}
\centering
\includegraphics[width=6 cm]{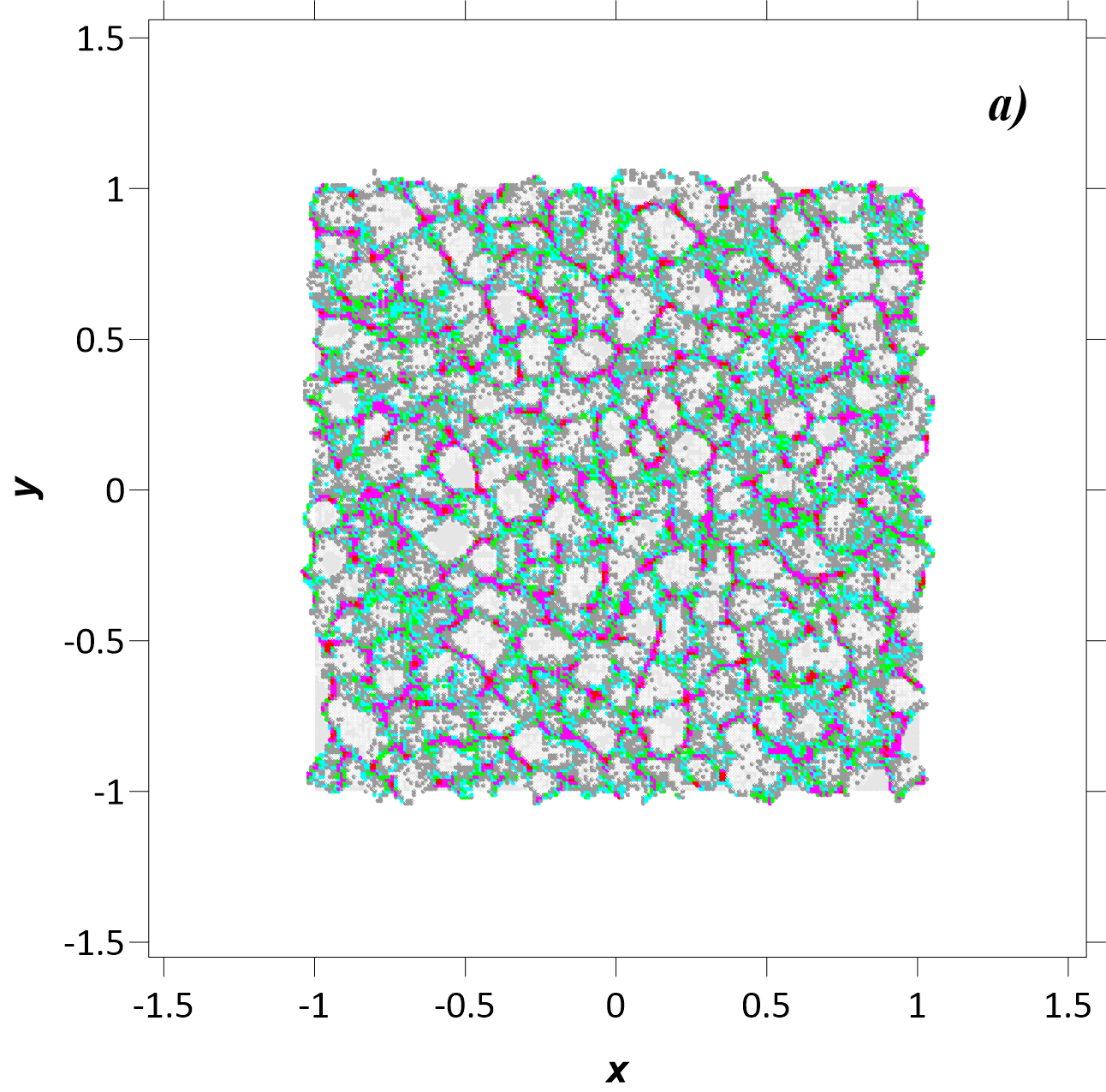}
\includegraphics[width=6 cm]{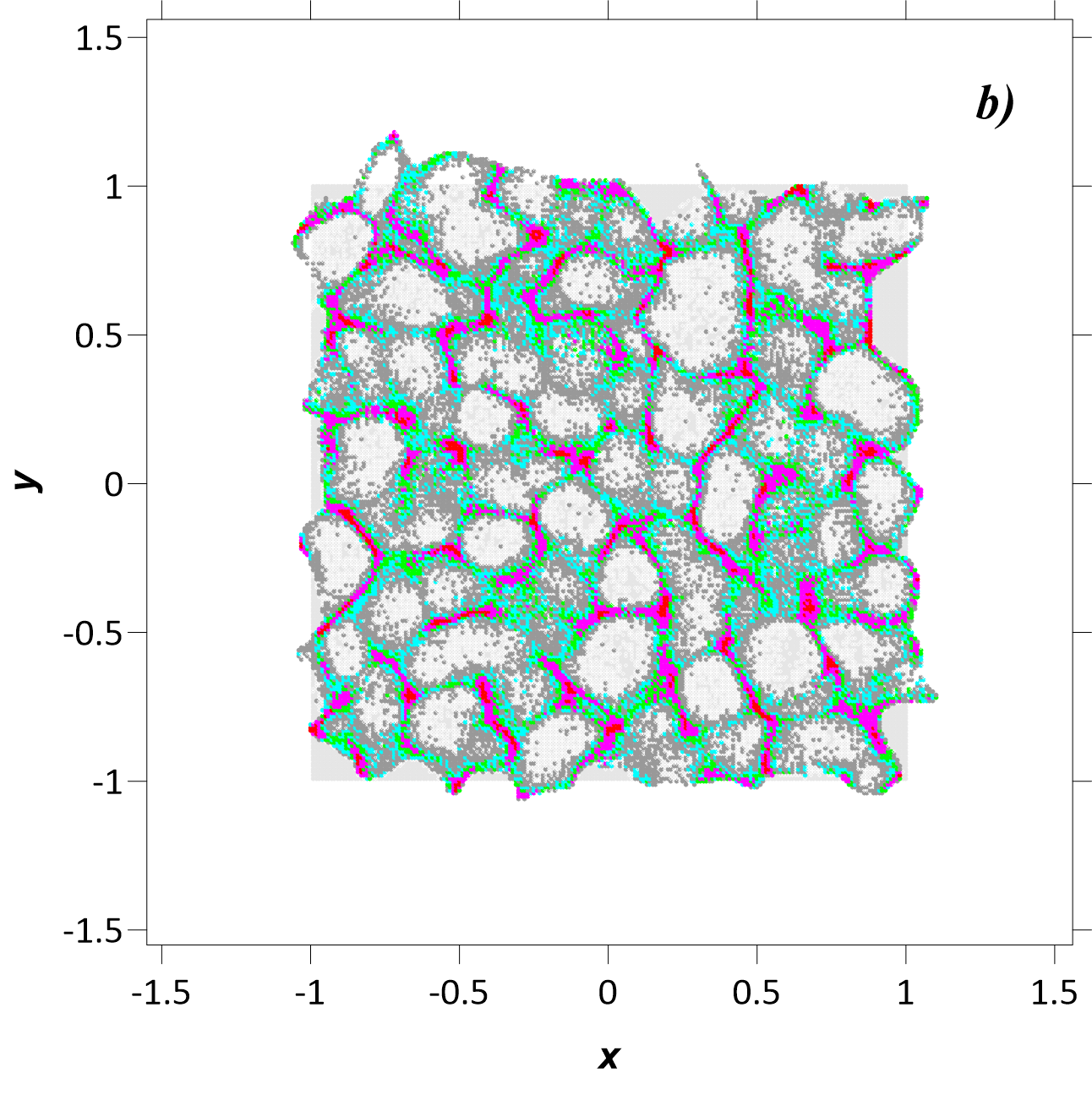}
\includegraphics[width=6 cm]{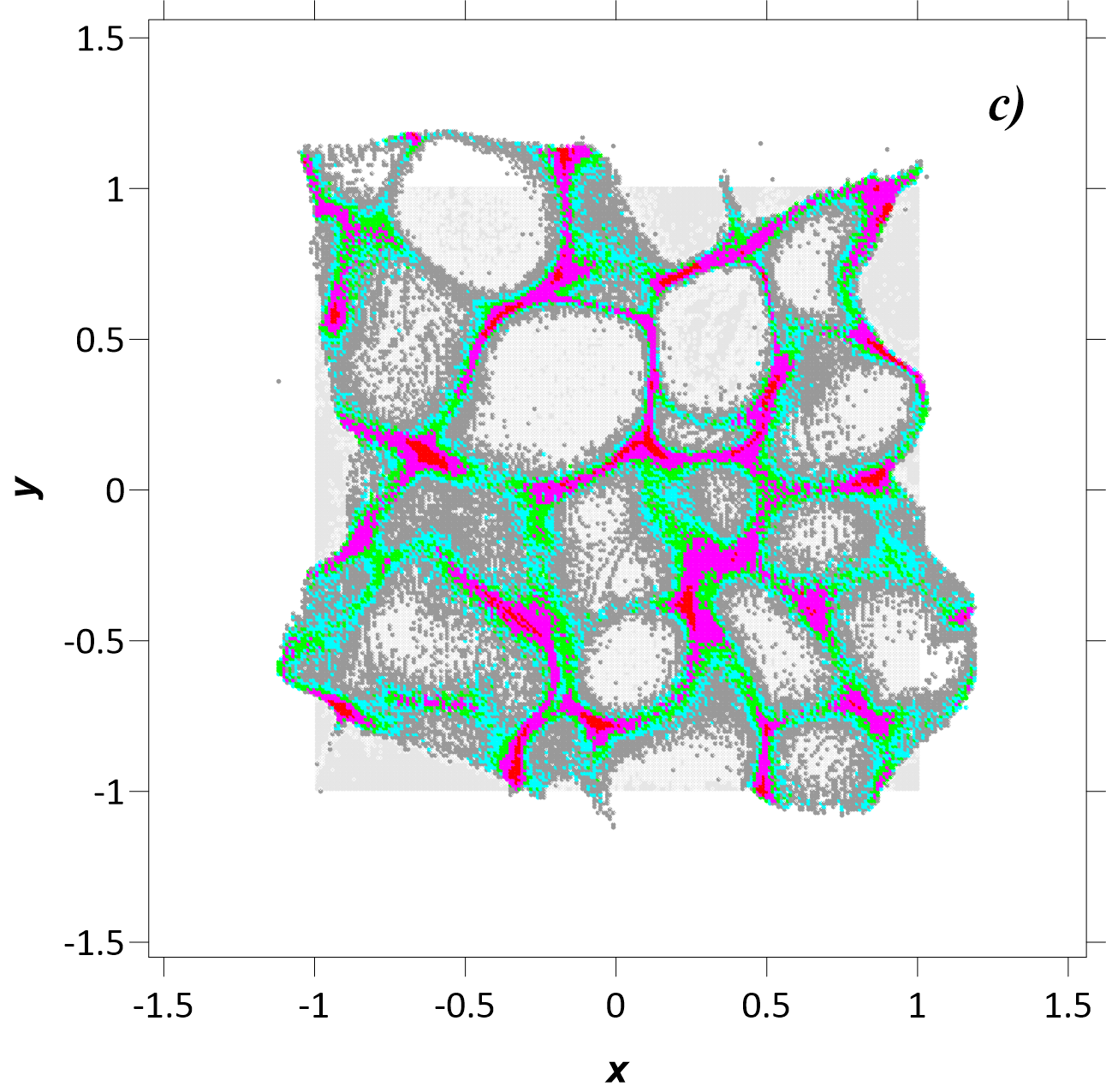}
\includegraphics[width=5 cm]{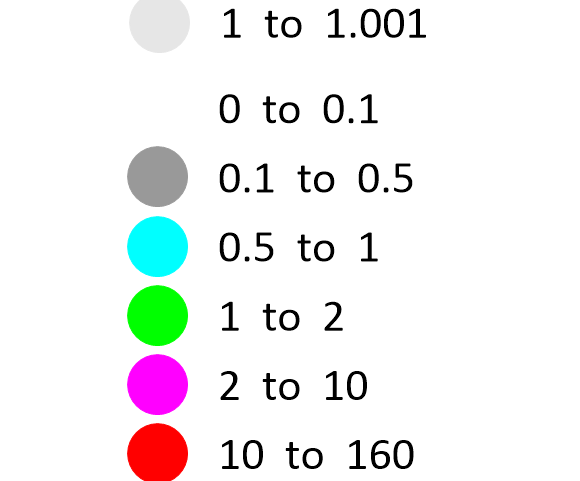}
\caption{
\label{fig2}
Same as Fig. \ref{fig1} but for the concentration distribution.
}
\vskip20pt
\end{figure}

\begin{figure}
\centering
\includegraphics[width=6 cm]{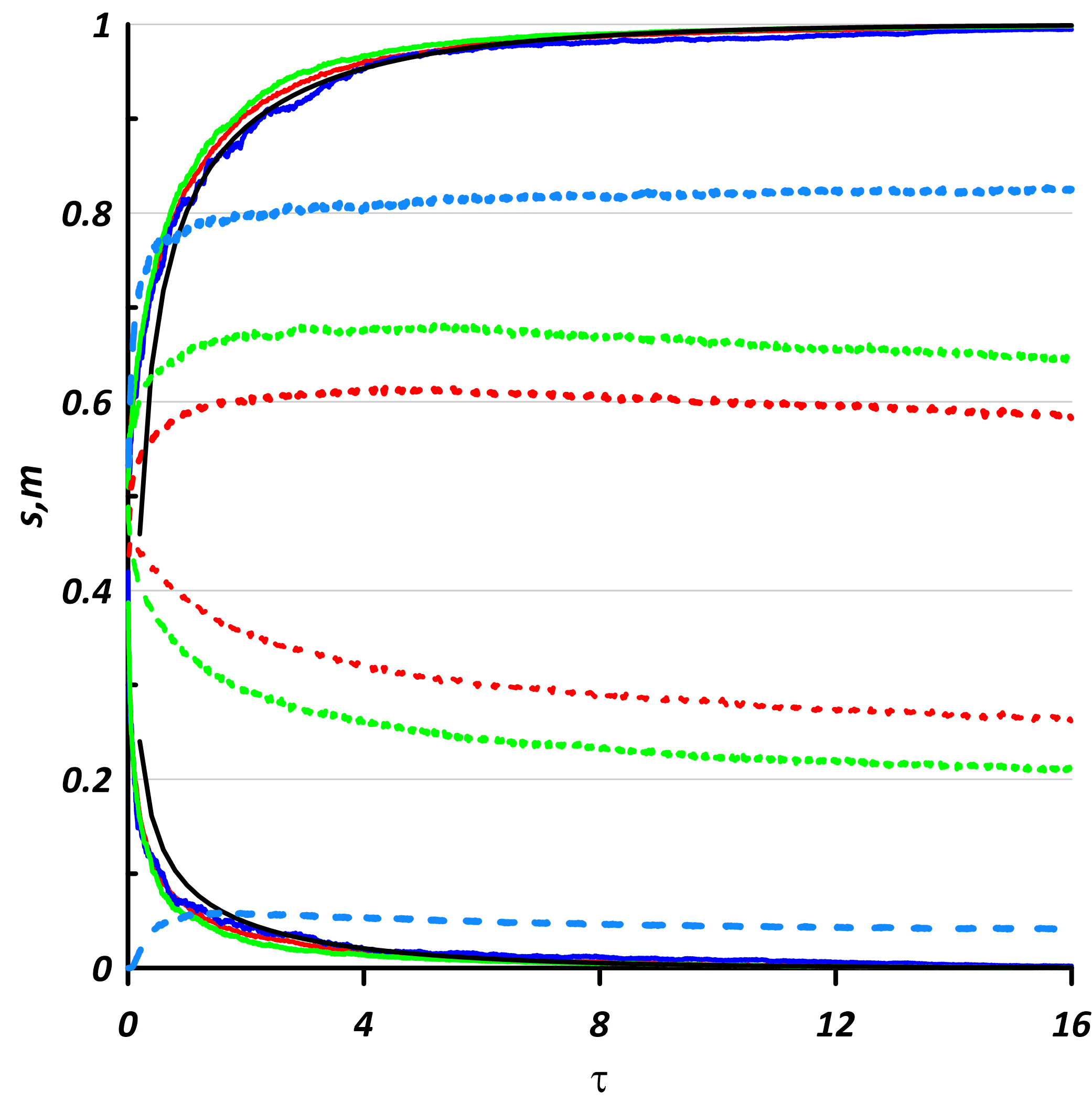}
\caption{
\label{fig3}
Time evolution of the clustering area (decreasing curves) and mass (increasing curves) corresponding to the
threshold value $\bar\rho=1$ and a single velocity realization.
Parameters correspond to Fig. \ref{fig1}: (a) red, (b) green, (c) blue lines.
The black lines represent the asymptotics (\ref{log_sm}).
The dashed lines illustrate the corresponding solution for $\gamma=0$ (Fig. \ref{fig4}).
}
\vskip20pt
\end{figure}

First, let us consider the limiting case of purely divergent velocity ($\gamma=1$). 
Evolution of particles is computed over the diffusion time scale $1/D^p$ and for different values of the spatial
correlation radius $l_{corr}$ (Fig. \ref{fig1},\ref{fig2}).
Clustering is evident in the concentration field (Fig. \ref{fig2}), but it is more pronounced in the density field (Fig. \ref{fig1}).
Fig. \ref{fig3} shows that the clustering area and mass are very close to the asymptotic exponential prediction, even in individual realizations.
We also found that the clustering depends weakly on the spatial correlation radius, provided the time is scaled
diffusively. The dependency from spacial correlation radius has two aspects. One is the the speed of clustering by virtue of diffusion time scale \ref{num_dsp}) which depend from value of the correlation radius. Second one is the space structure of the clusters. We can see qualitatively that particles aggregates in thin lines and the distances between these lines are proportional to spacious correlation radius. 
%***EXPLAIN HERE HOW EXACTLY IT DEPENDS***

\begin{figure}
\centering
\includegraphics[width=6 cm]{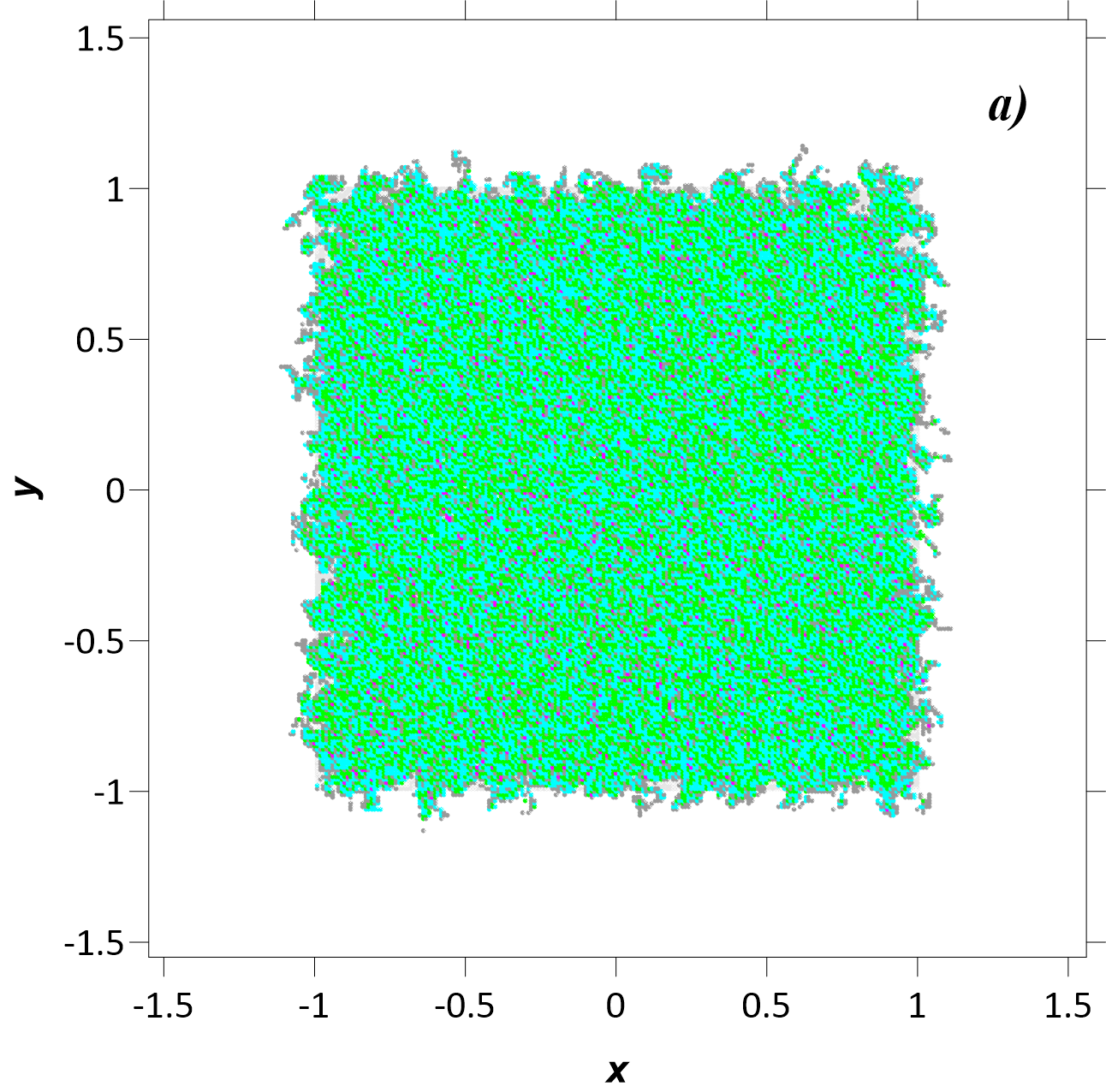}
\includegraphics[width=6 cm]{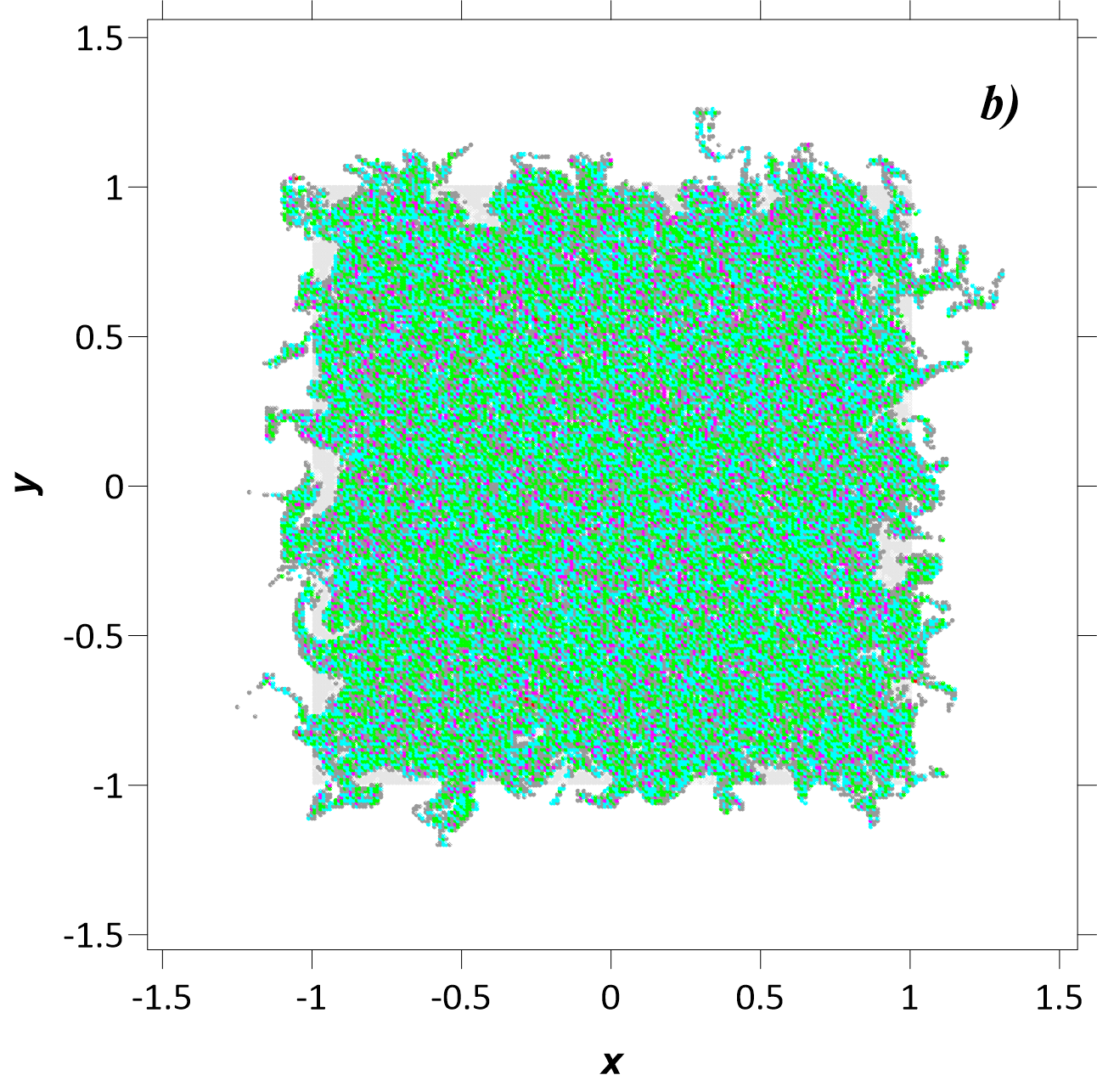}
\includegraphics[width=6 cm]{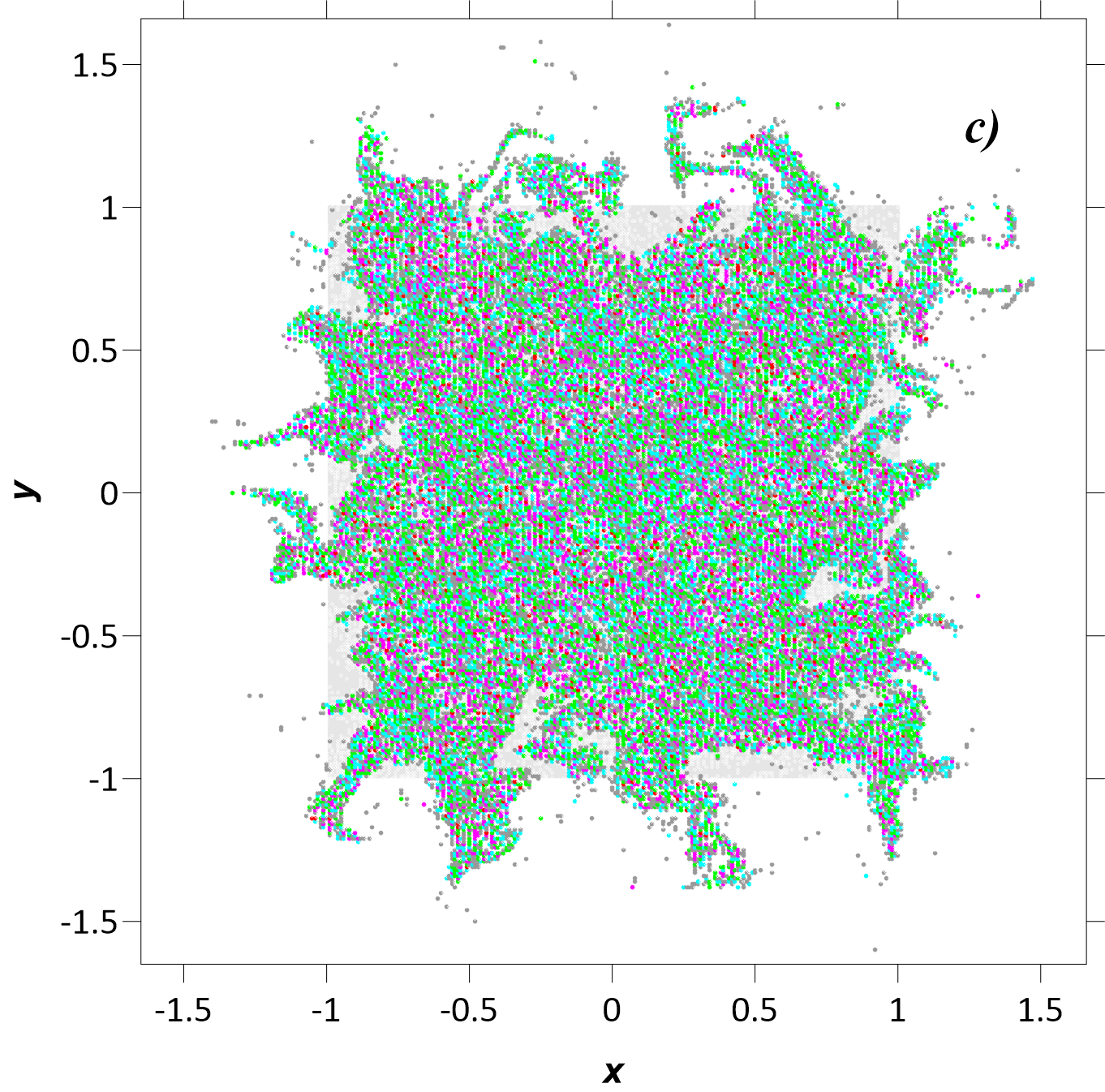}
\includegraphics[width=4 cm]{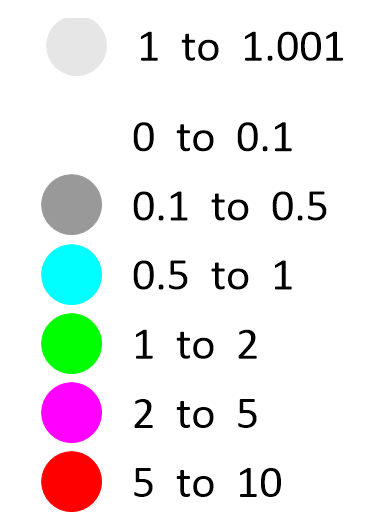}
\caption{
\label{fig4}
Same as Fig. \ref{fig1} but for the purely solenoidal velocity ($\gamma=0$).
}
\vskip20pt
\end{figure}

\begin{figure}
\centering
\includegraphics[width=5.5 cm]{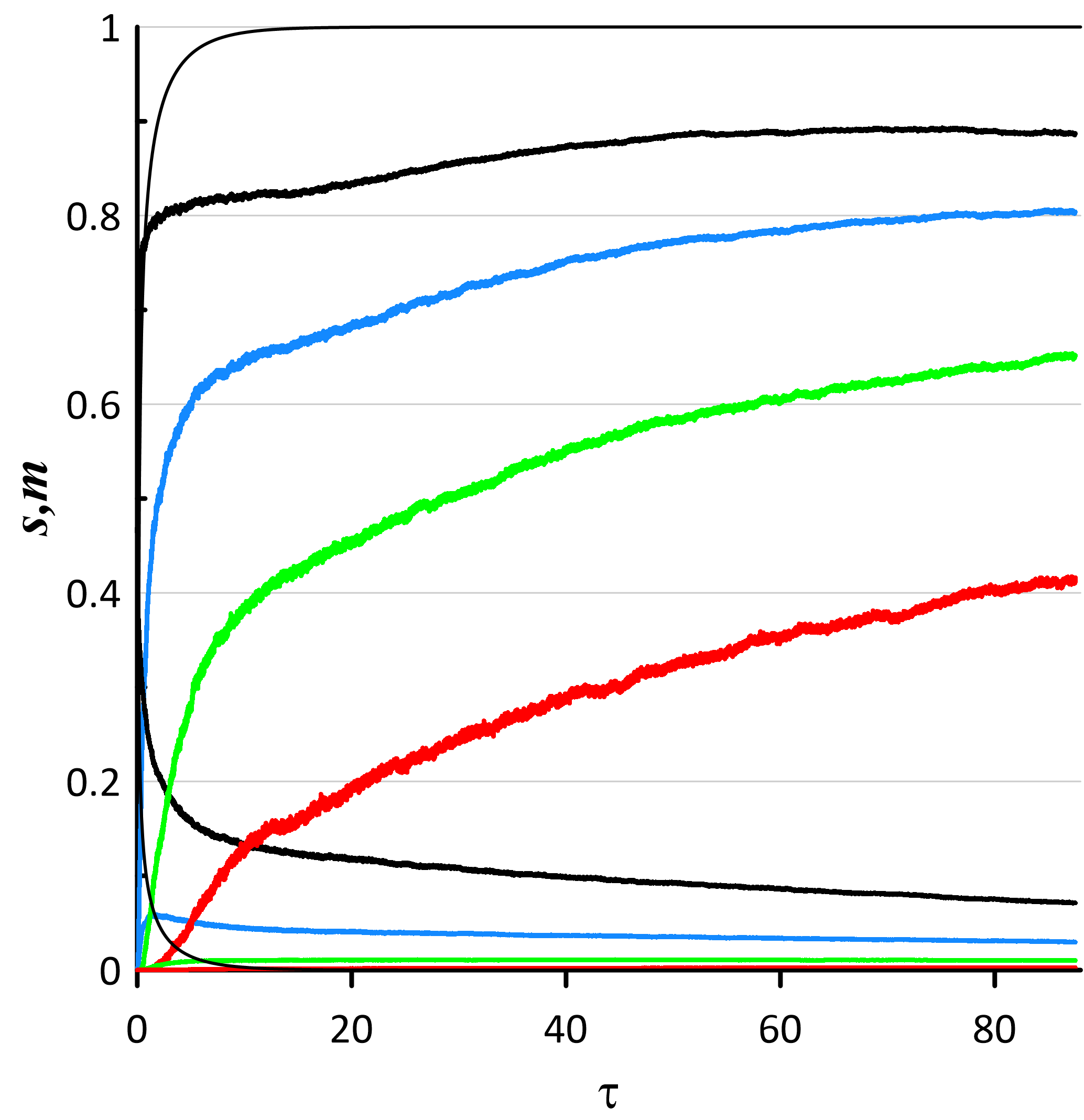}
\caption{
\label{fig5}
The clustering area and mass for the purely nondivergent velocity regime ($\gamma=0$) and for a single flow realization.
Parameters are as in Fig. \ref{fig4}c and the curves correspond to: $\bar\rho=1$ --- black; $\bar\rho=2$ --- blue;
$\bar\rho=4$ --- green; $\bar\rho=8$ --- red colors.
The thin black curve corresponds to the asymptotic theory (\ref{log_sm}) and clearly shows the diffusion time scale.
}
\vskip20pt
\end{figure}

\begin{figure}
\centering
\includegraphics[width=6 cm]{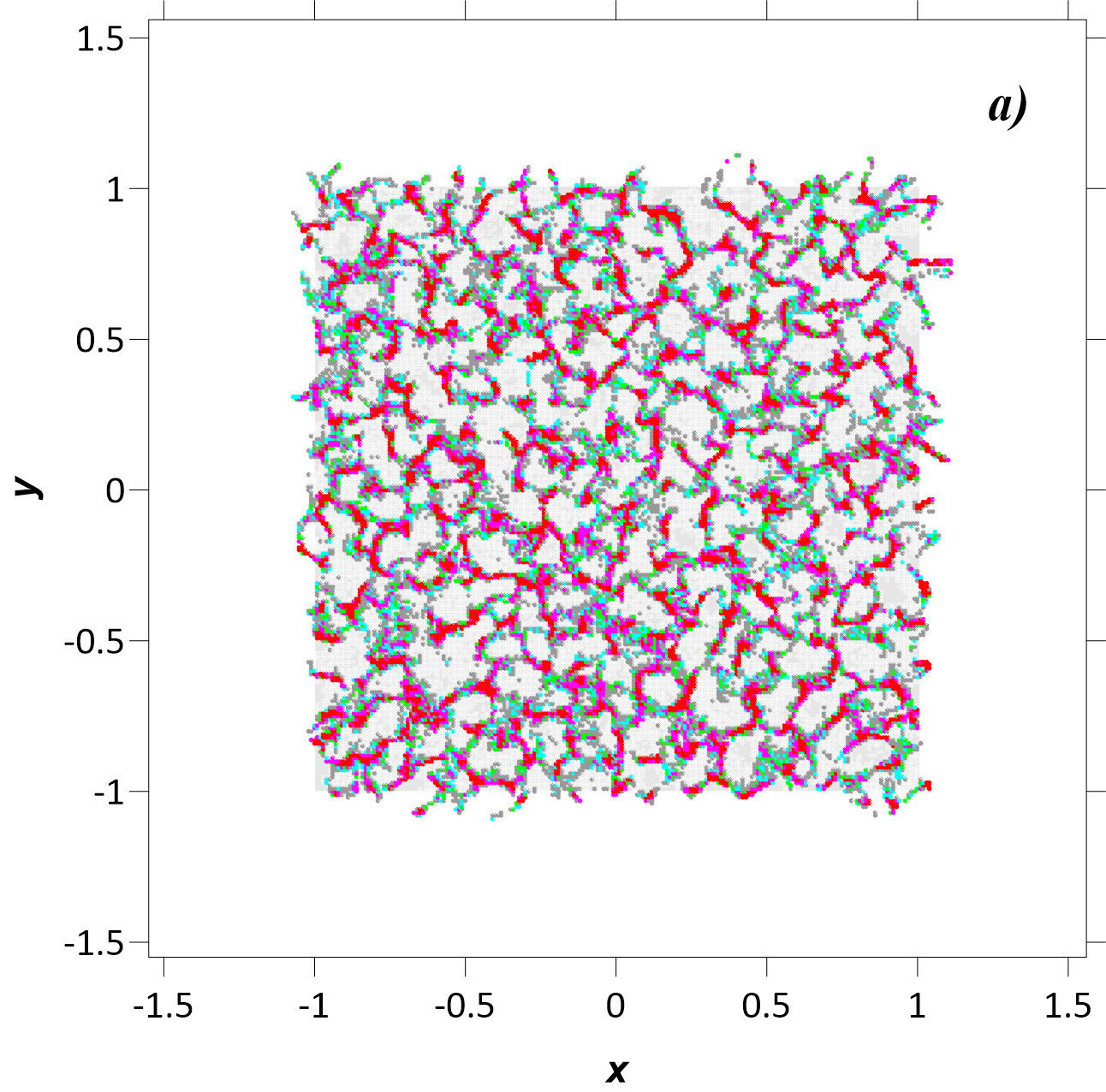}
\includegraphics[width=6 cm]{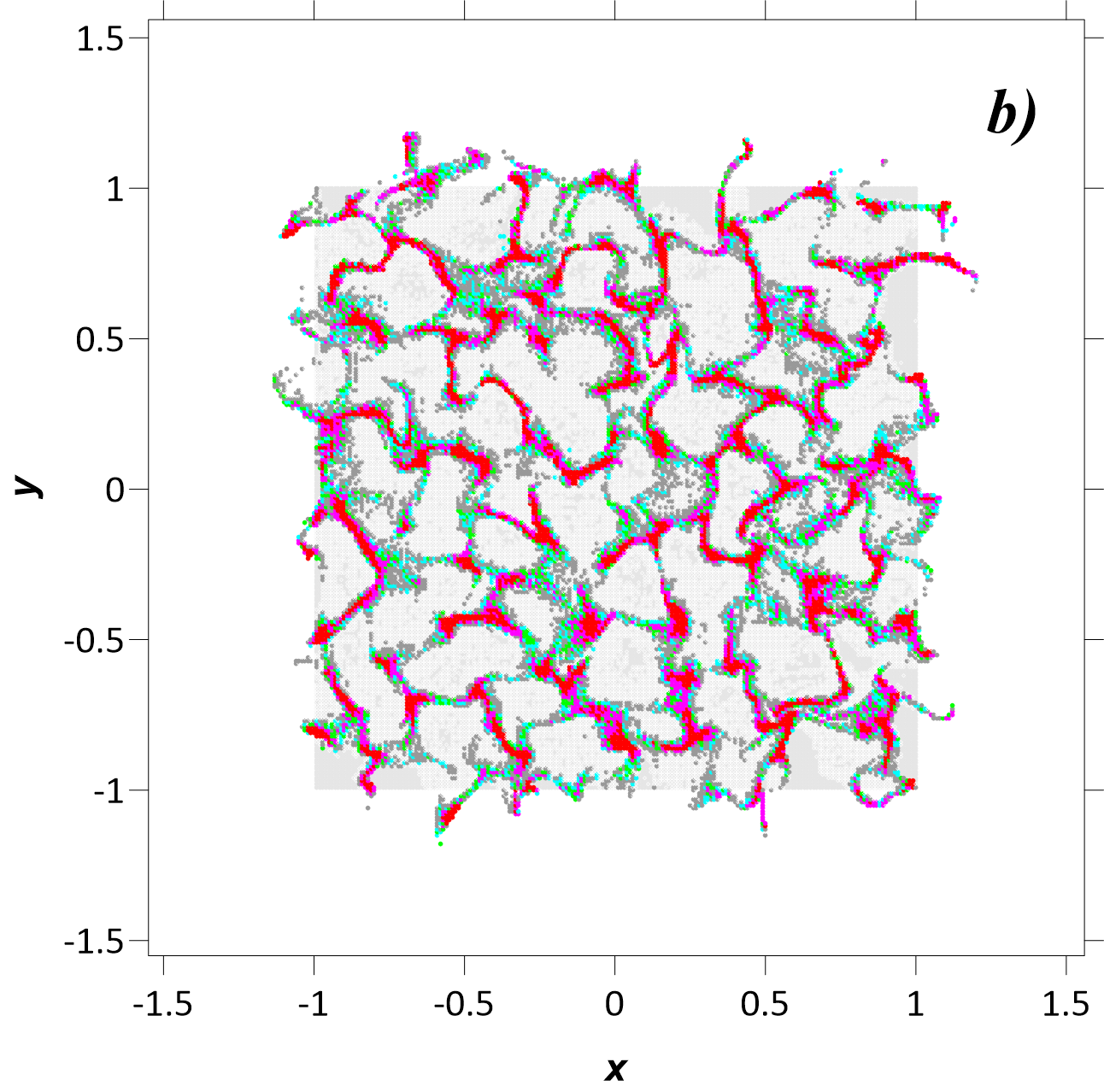}
\includegraphics[width=6 cm]{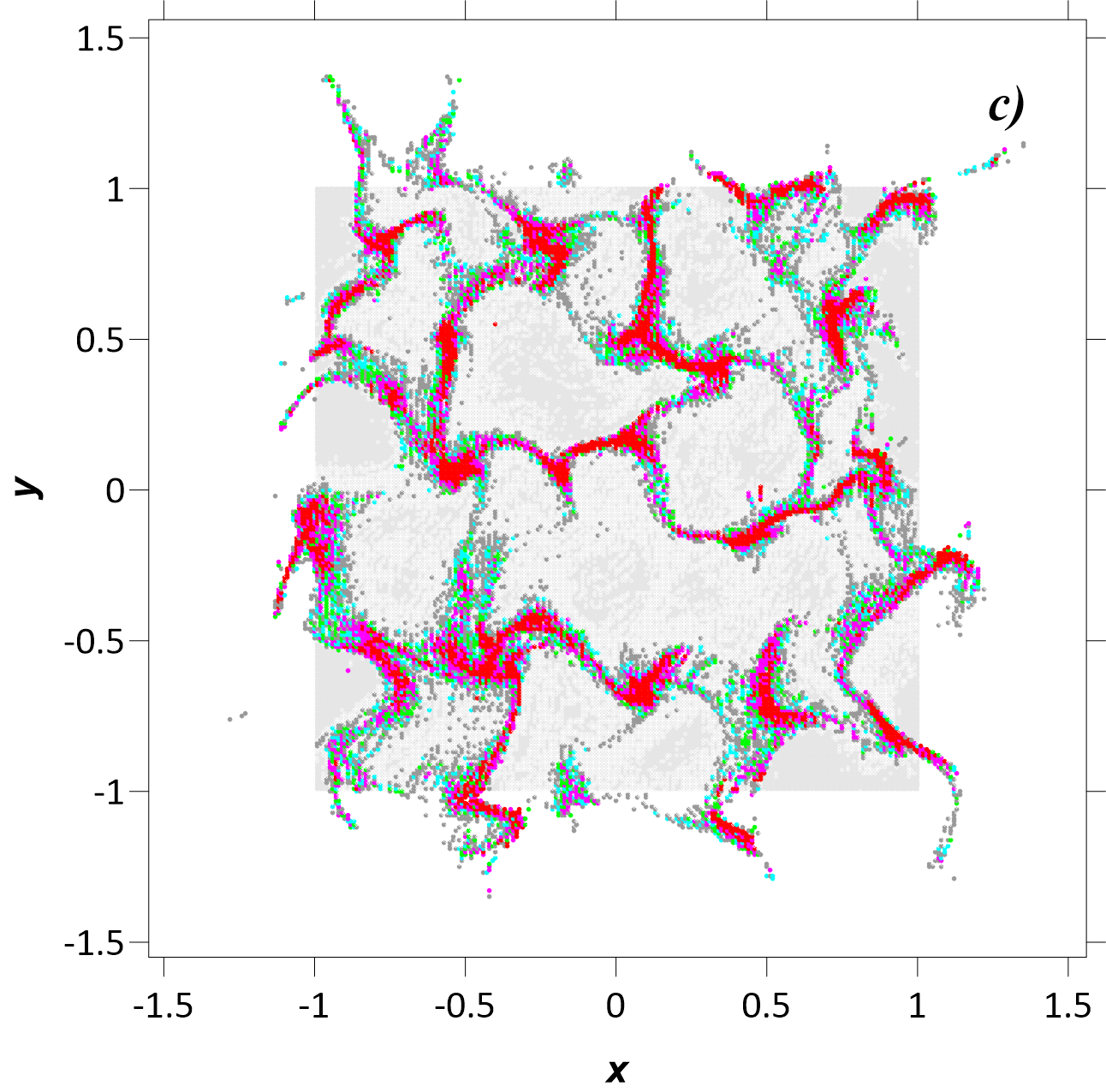}
\includegraphics[width=4.0 cm]{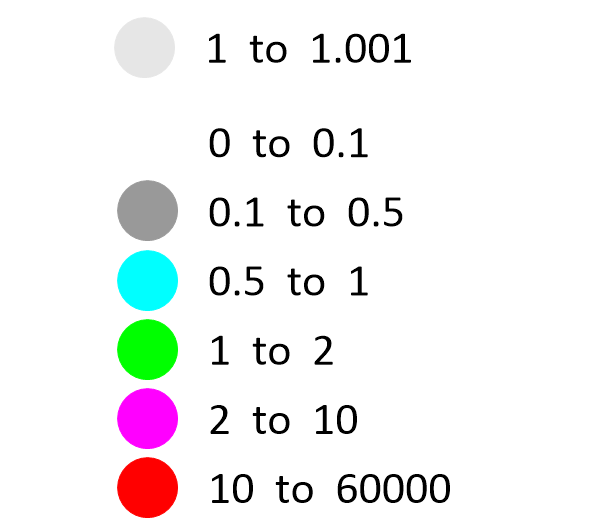}
\caption{
\label{fig6}
Density distribution for $\gamma=0.5$ and $\tau=4$, which is the divergent diffusion time:
(a) $l = 0.04, \hskip5pt \sigma_U=0.33$;
(b) $l = 0.08, \hskip5pt \sigma_U= 0.67$;
(c) $l = 0.16, \hskip5pt \sigma_U= 0.67$.
The rest is as in Fig. \ref{fig1}.
}
\vskip20pt
\end{figure}

Let us now compare clustering in the purely divergent ($\gamma=1$) and non-divergent ($\gamma=0$) flow regimes
(Figs. \ref{fig4} and \ref{fig5}).
In the latter regime there is no difference between the density and concentration clustering, because the
density is also materially conserved.
Furthermore, in this regime the asymptotic theory (\ref{tip_eq}) predicts no complete clustering at all.
Instead, both fields are affected by the stirring; as a result, they develop clusters due to the fragmentation
process, which is fundamentally different from the complete clustering process.
{\er These clusters are transient and never aggregate large values of density and concentration. This is because the stirring process reshuffles the particles at each time instant because of the kinematic nature of the random velocity field. 
Presence of dynamical constraints in the velocity field might in fact significantly alter this type of clustering leading to anomalously large values provided the velocity field is induced by coherent structures such as vortices and jets. Investigating this possibility is beyond the scope of the paper and will be addressed elsewhere.}
As a result, the density excess over threshold 
%***WHY DOES IT HAPPEN?!*** 
value is moderate: the number of grid cells with $\rho({\bf R},\tau) > 2$ is small.
Moreover, the maximum density value (about 5 after $t=3\tau$) is significantly smaller than in the purely
divergent case (where it is more than 100 at $t=\tau$). 
Strictly speaking, these high density (or concentration) values do not signify the complete clustering
in the sense of estimates (\ref{tip_eq}) and (\ref{length}), but reflect the fragmentation clustering.
Fig. \ref{fig3},\ref{fig5} shows the clustering area and mass based on the density larger than $\bar\rho=1$.
The clustering mass decreases, but less so for a large correlation radius ($l_{corr}=0.16$), because of the
slowed down fragmentation process.
The clusters are typically small and uniformly distributed, and there are no clusters on the scale of $l_{corr}$. 
We hypothesize that fragmentation clustering can dominate over exponential clustering in the presence of 
diffusion ($\kappa \ne 0$), but this analysis is beyond the scope of the paper.

\begin{figure}
\centering
\includegraphics[width=6 cm]{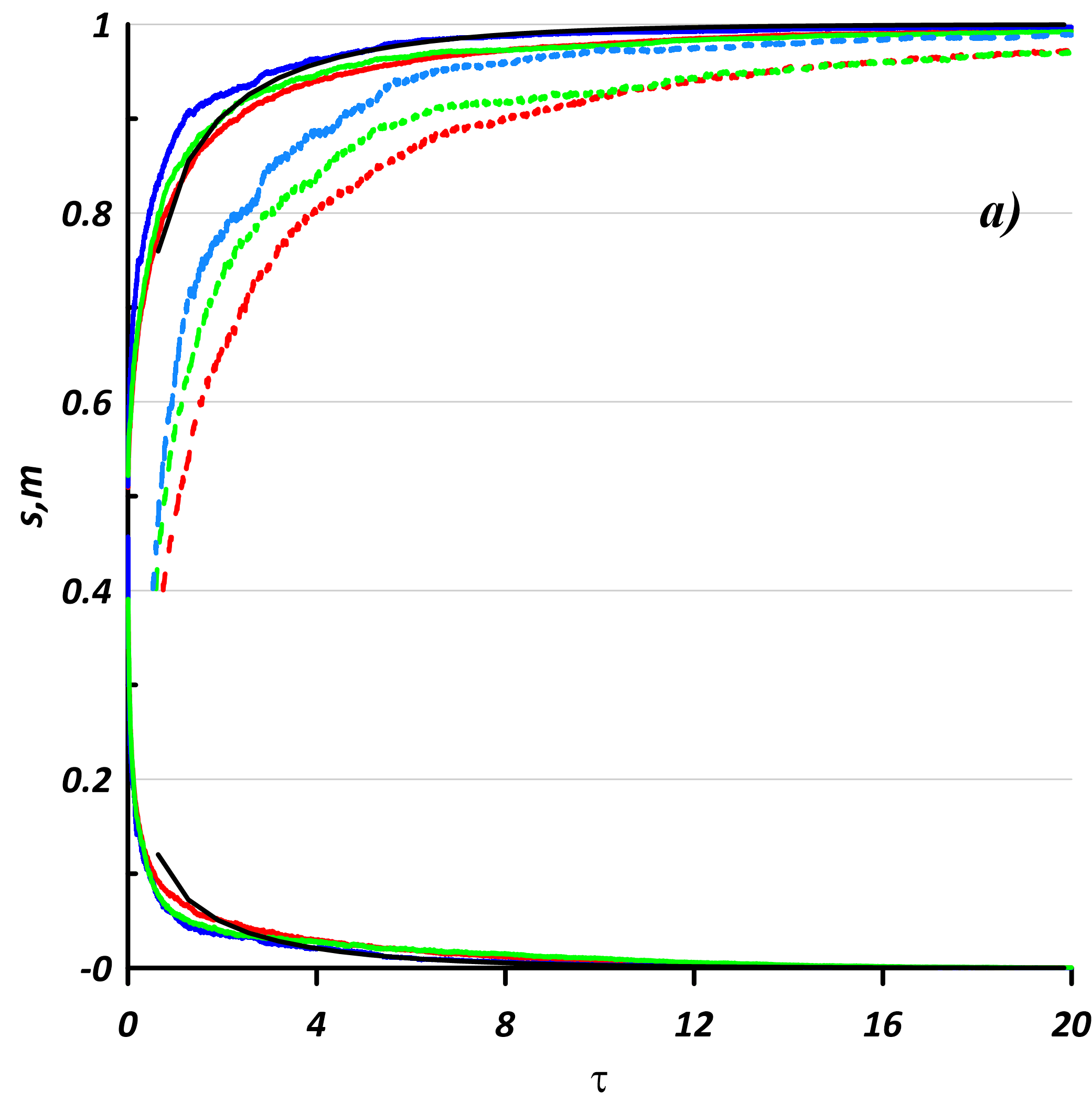}
\includegraphics[width=6 cm]{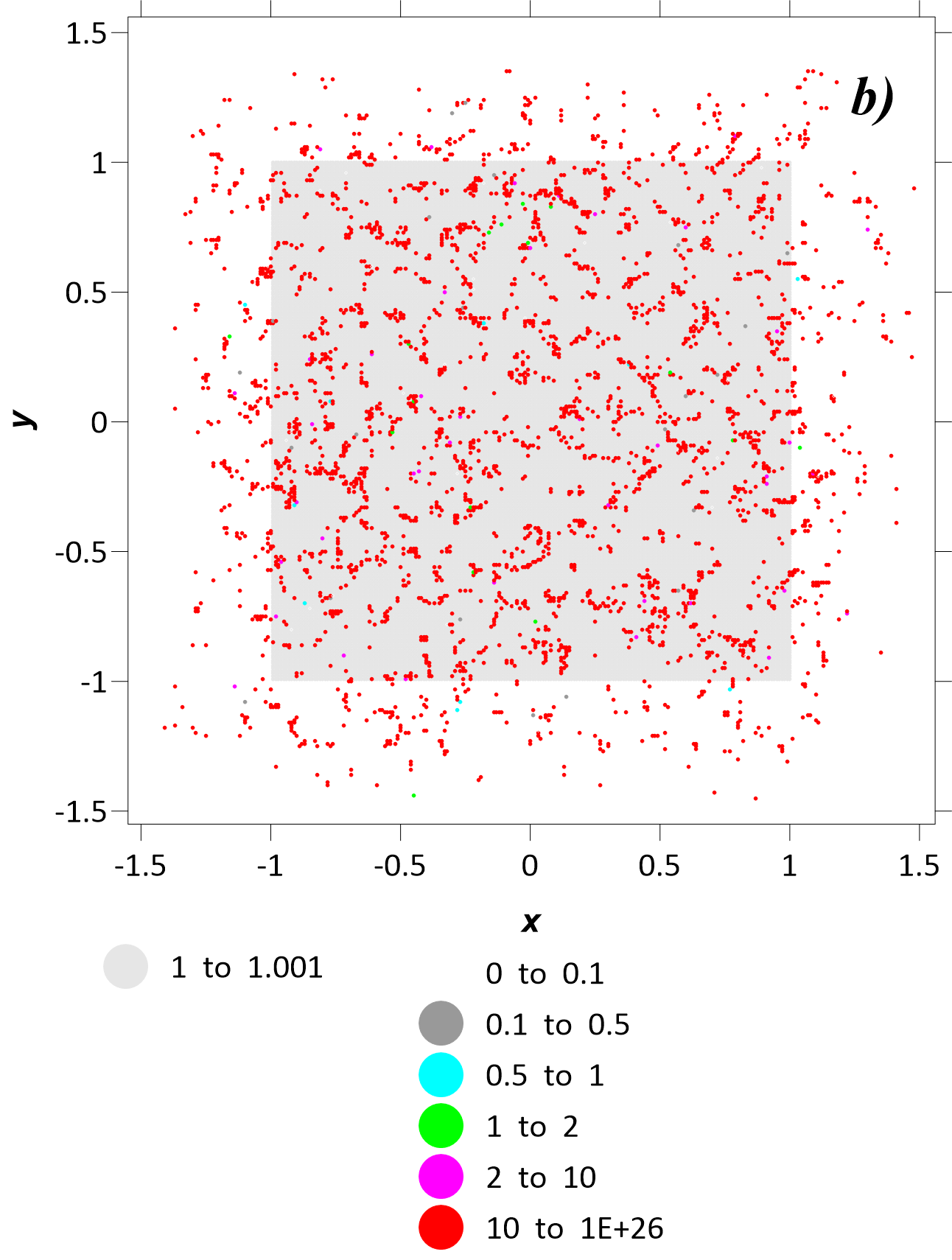}
\caption{
\label{fig7}
Evolution of the clustering area and mass ($\bar\rho=1$) for a single realization and $\gamma=0.5$, for:
(a) $l = 0.04, \hskip5pt \sigma_U=0.33$ --- red; $l = 0.08, \hskip5pt \sigma_U=0.67$ --- green;
$l = 0.16, \hskip5pt \sigma_U=0.67$ --- blue curves.
The black lines represent the asymptotics.
The dashed lines correspond to the threshold value $\bar\rho=4$.
(b) Density distribution for $l = 0.04, \hskip5pt \sigma_U=0.33$ and time $t=18\tau$.
The rest is as in Fig. \ref{fig6}.
}
\vskip20pt
\end{figure}

\begin{figure}
\centering
\includegraphics[width=6 cm]{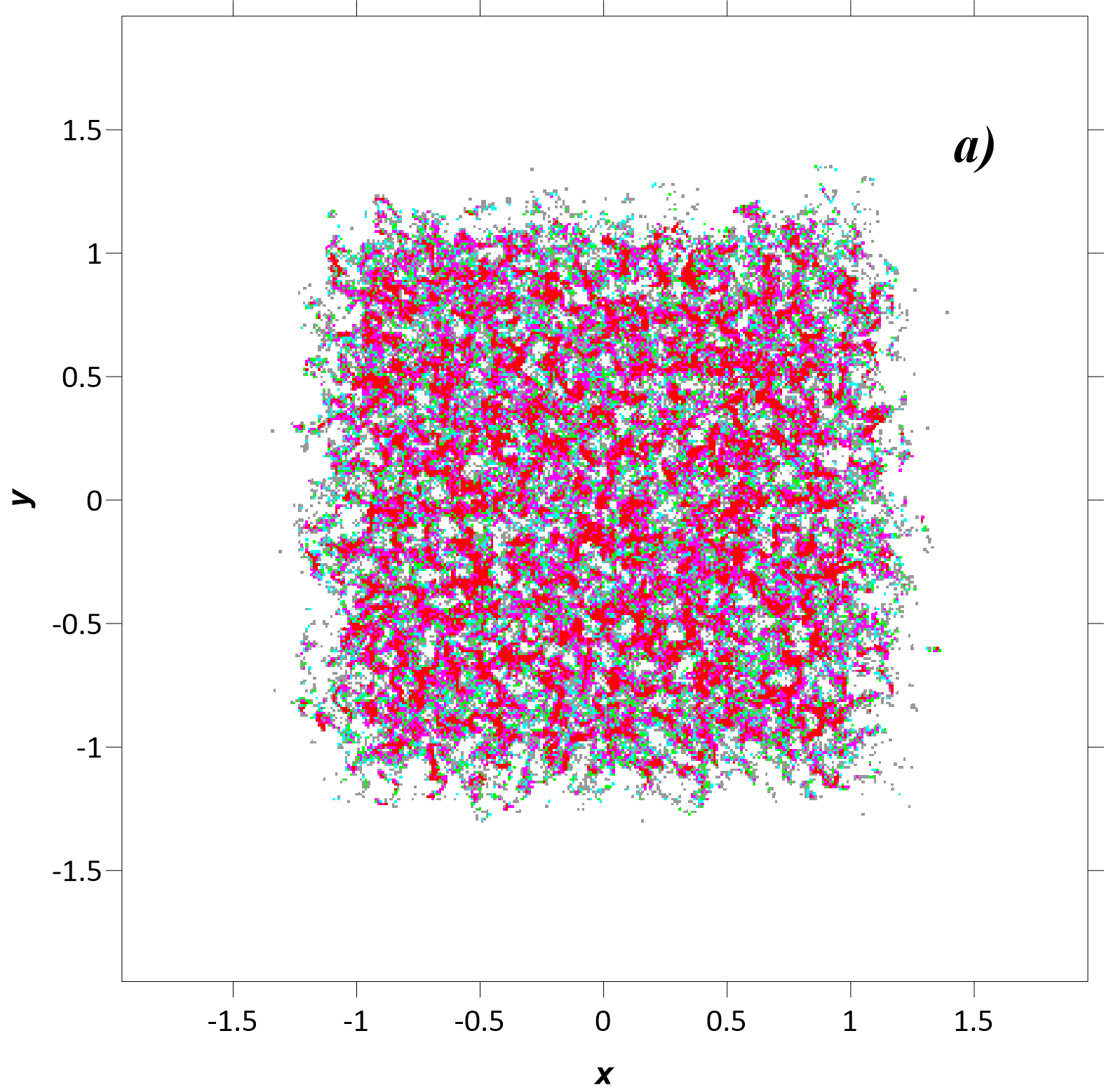}
\includegraphics[width=6 cm]{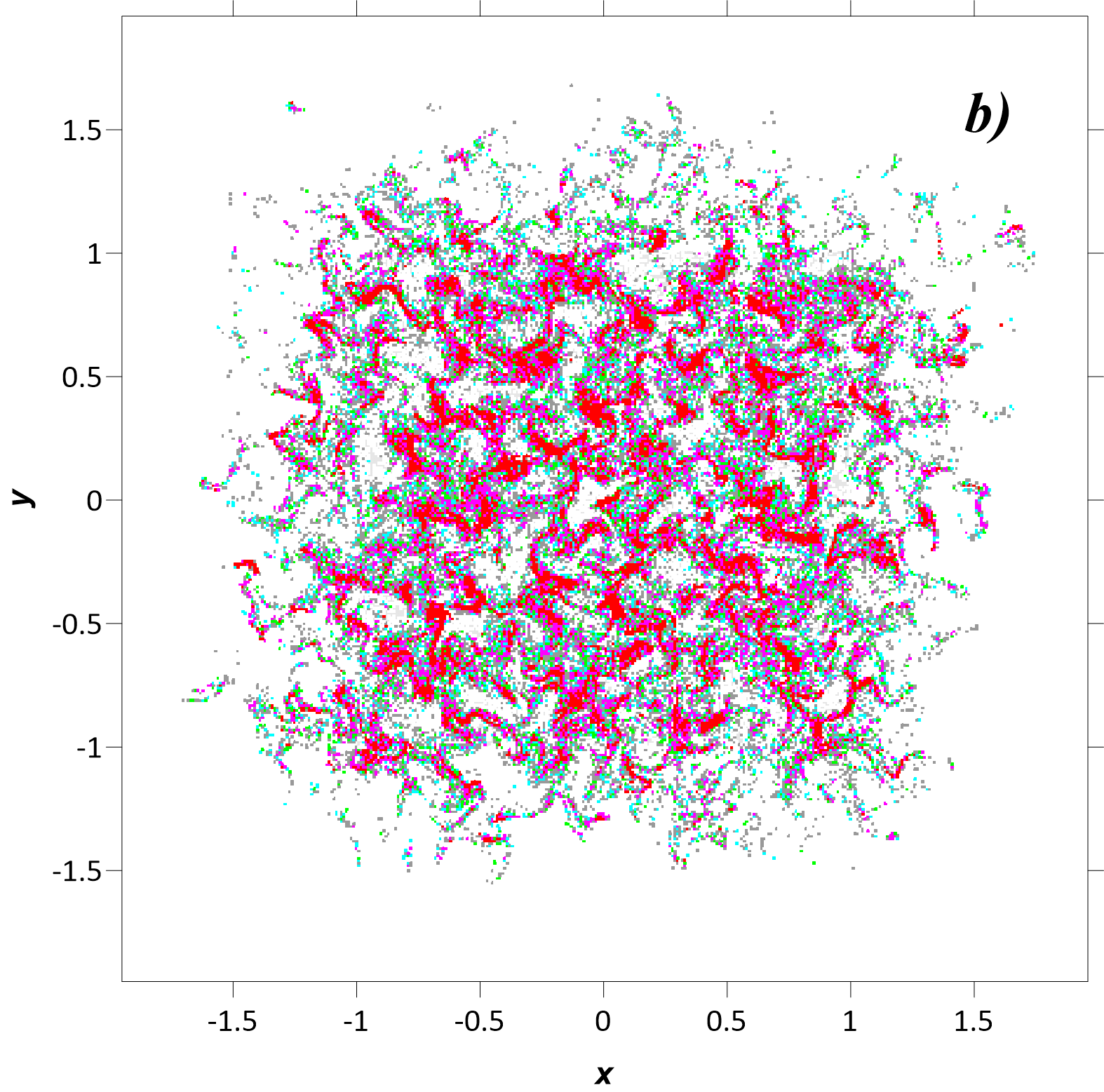}
\includegraphics[width=7.8 cm]{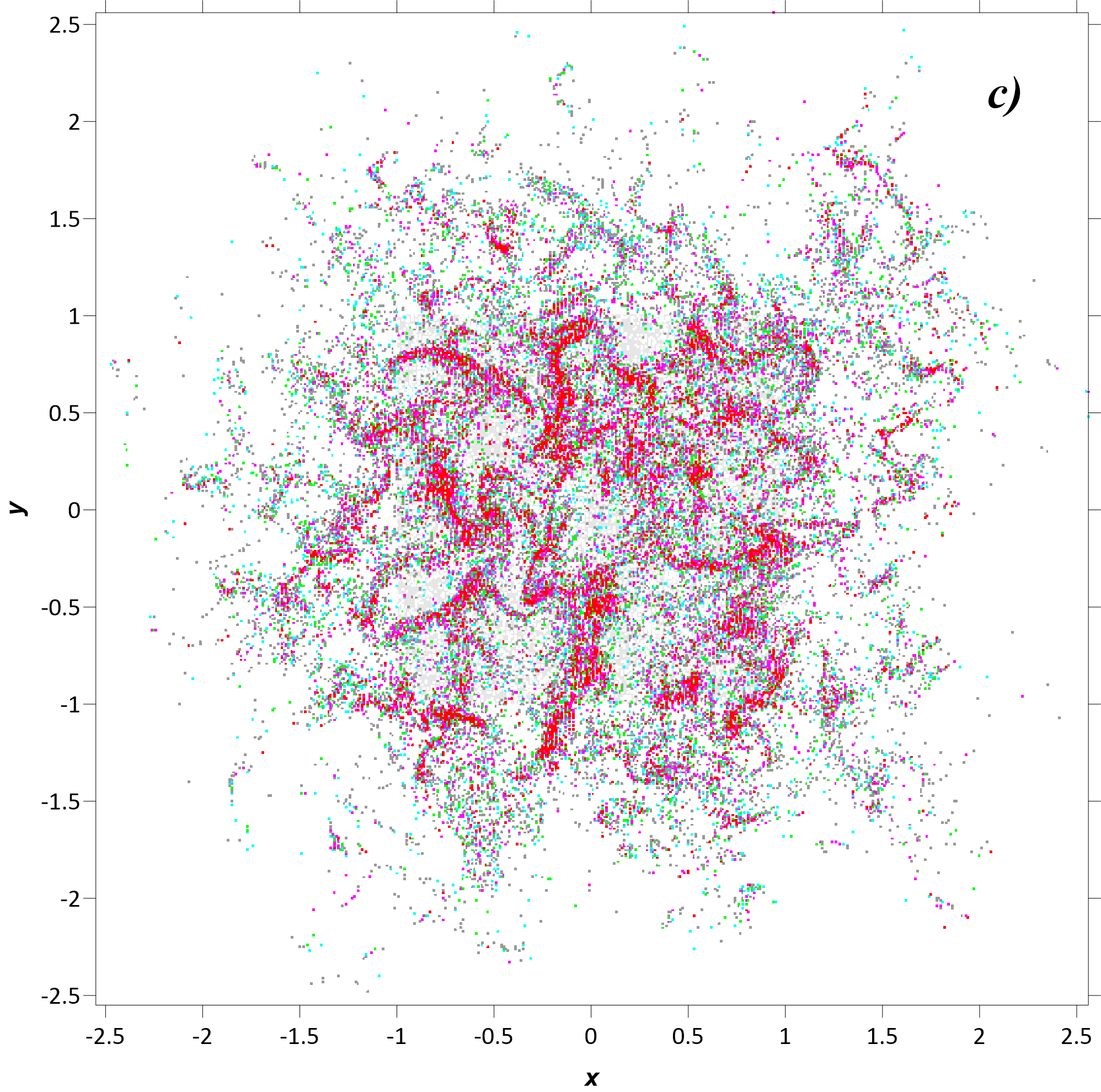}
\includegraphics[width=4.0 cm]{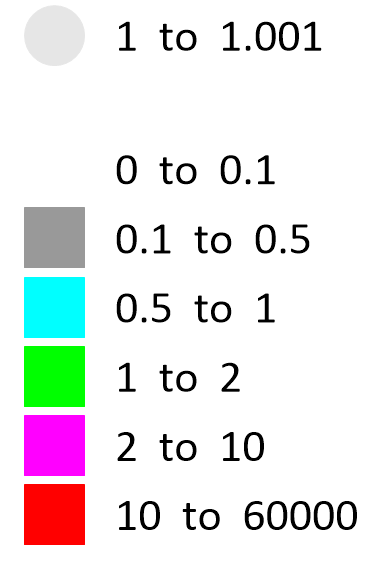}
\caption{
\label{fig8}
Density distributions for $\gamma=0.2$, $t=25\tau$:
(a) $l = 0.04, \hskip5pt \sigma_U=0.33$;
(b) $l = 0.08, \hskip5pt \sigma_U=0.67$;
(c) $l = 0.16, \hskip5pt \sigma_U=1.33$.
The rest is as in Fig. \ref{fig6}.
}
\vskip20pt
\end{figure}

%
%%%%%%%%%%%%%%%%%%%%%%%%%%%%%%%%%% WEAKLY DIVERGENT REGIME %%%%%%%%%%%%%%%%%%%%%%%%%%%%%%%%%%
%
\subsection{Weakly divergent regime}

Let us now consider a more realistic flow regime with equal contributions of the solenoidal (non-divergent)
and potential (divergent) components ($\gamma=0.5$; Fig. \ref{fig6}).
The asymptotic estimate (\ref{tip_eq}) still predicts no clustering; however, we have found that the clustering not
only occurs but is also very efficient.
Overall, the clusters develop at a much slower rate (Fig. \ref{fig7}), relative to the purely divergent flow
regime, but the complete clustering is eventually reached.
The clustering slowdown takes place for all $l_{corr}$ considered.
%, and it is more pronounced for larger density threshold (Fig. \ref{fig7}). ***WHY?***.
%As we can see above the clusters aggregate in thin lines in the purely divergent case ***THIS HAS BEEN NEVER
%DISCUSSED!***.
%Now the cluster lines become fragmented, which is due to the stirring induced by the solenoidal flow component.

\begin{figure}
\centering
\includegraphics[width=6 cm]{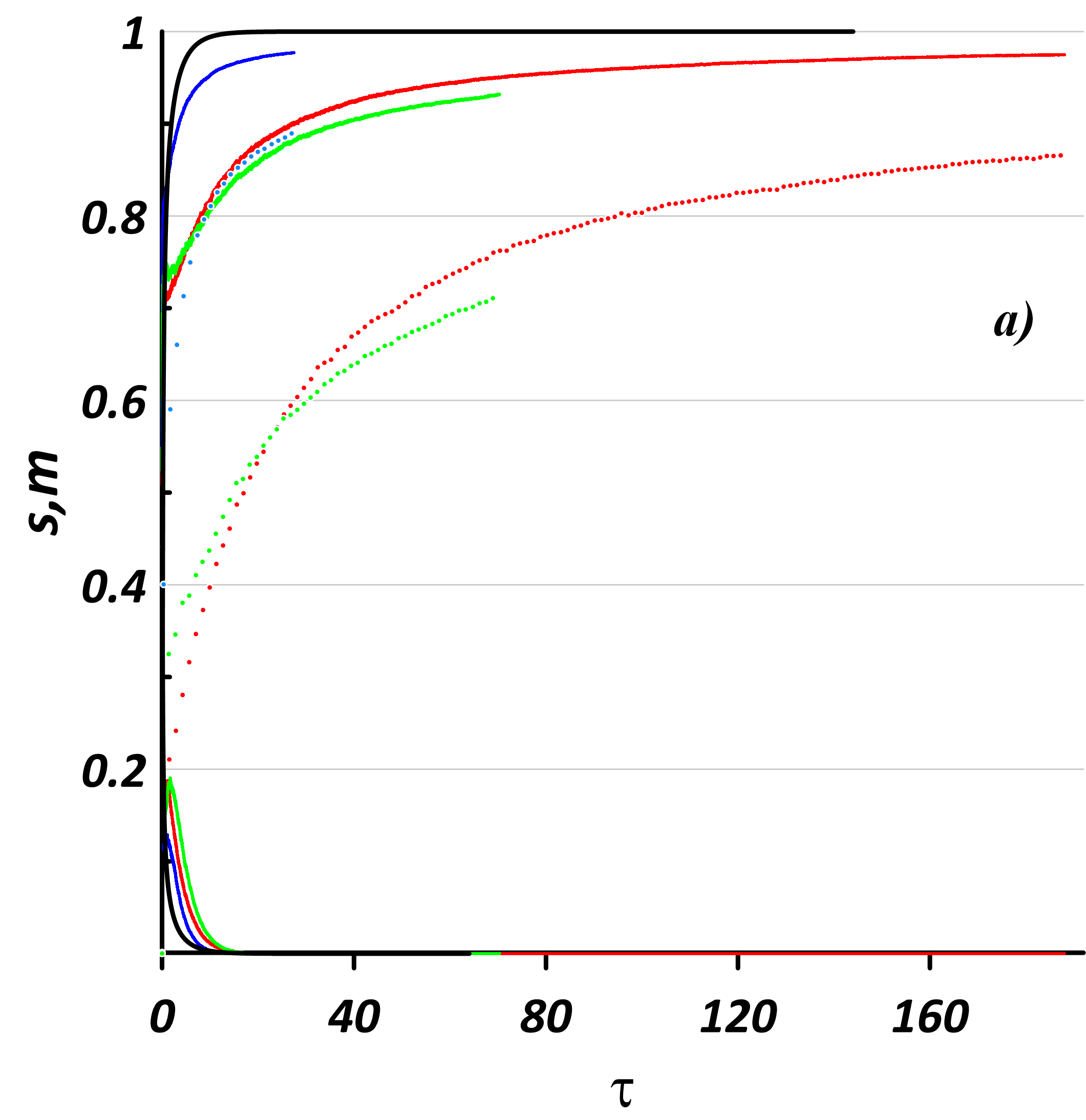}
\includegraphics[width=6 cm]{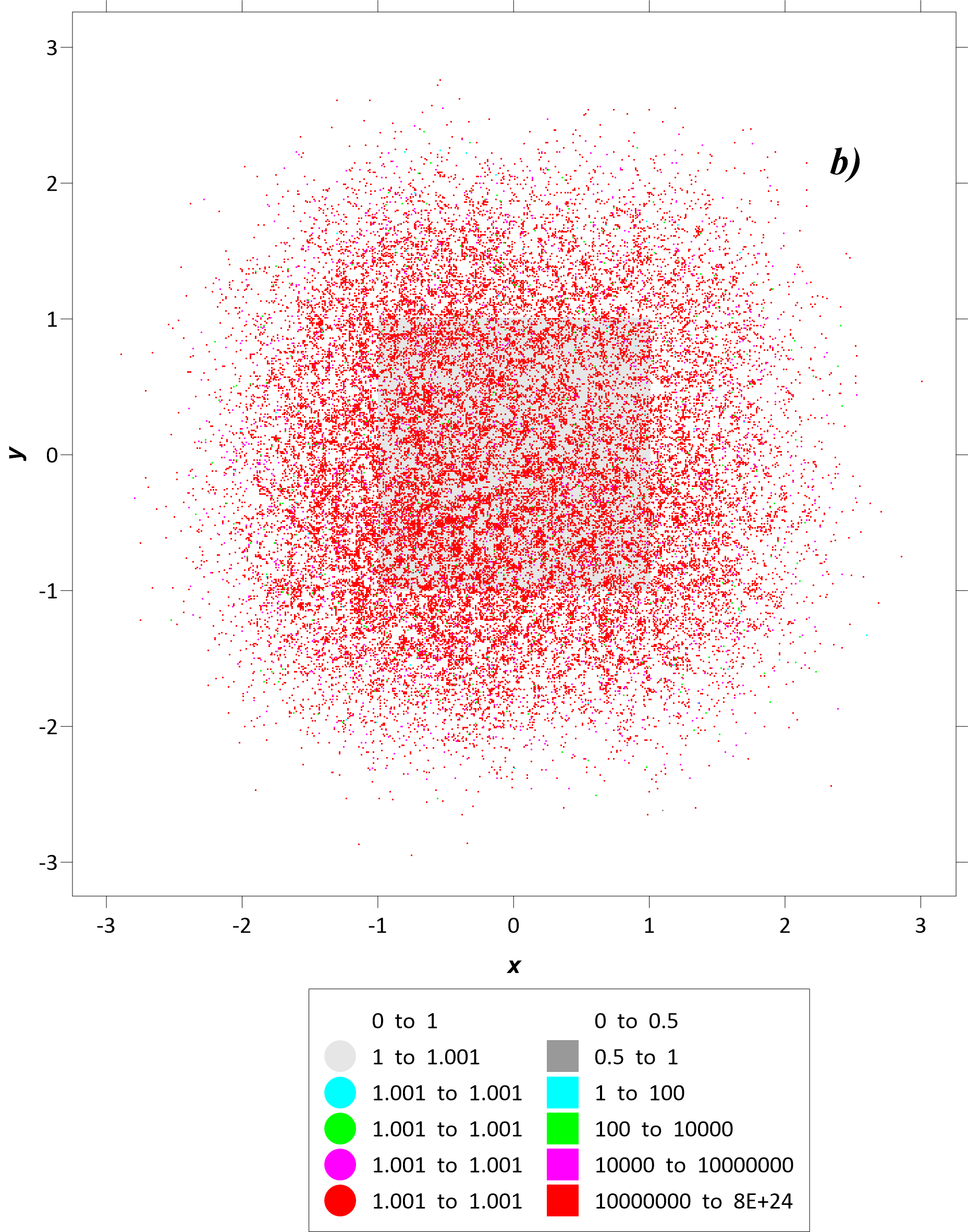}
\caption{
\label{fig9}
The same as Fig. \ref{fig7} but for $\gamma=0.2$.
}
\vskip20pt
\end{figure}

\begin{figure}
\centering
\includegraphics[width=6 cm]{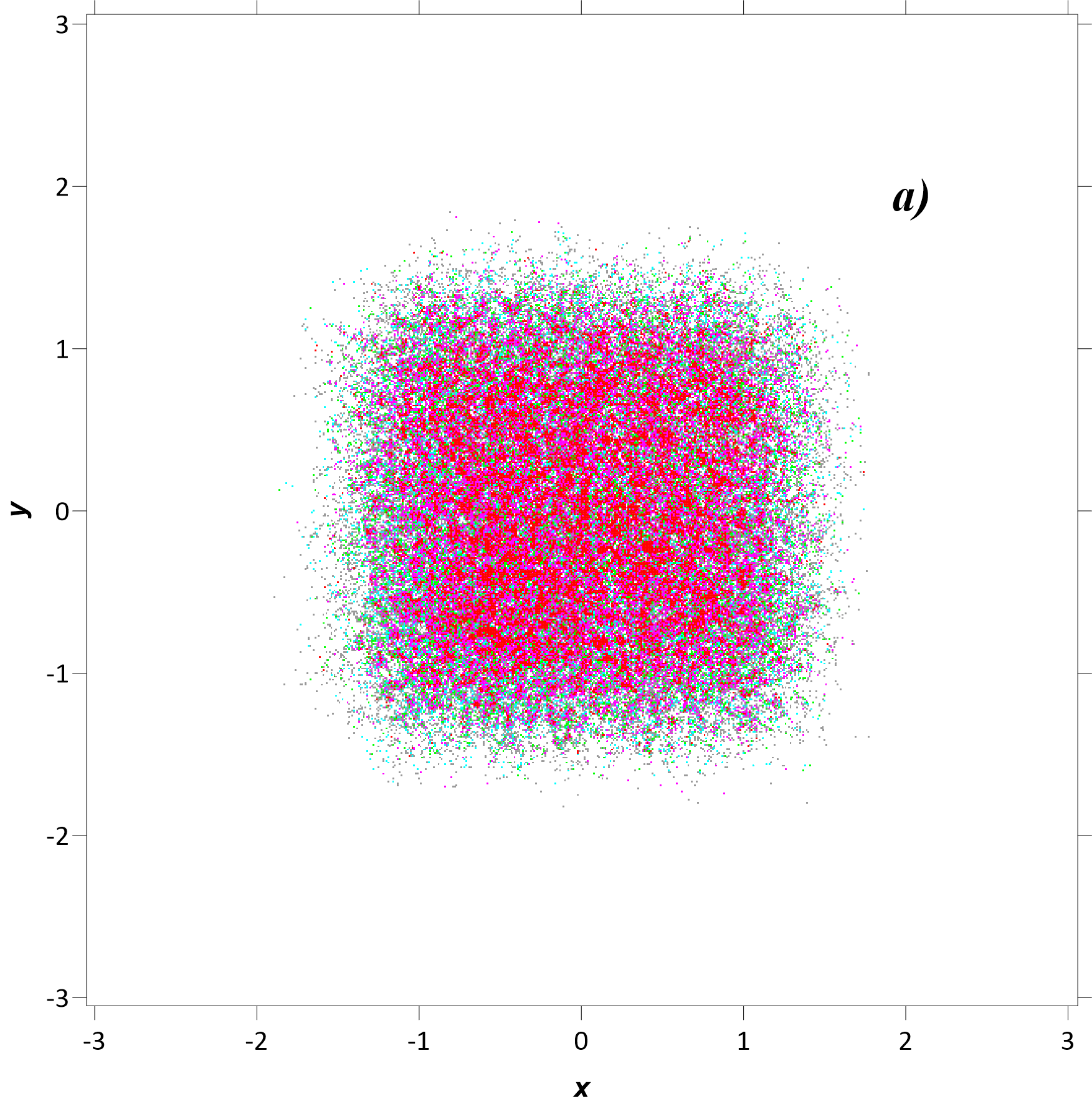}
\includegraphics[width=6 cm]{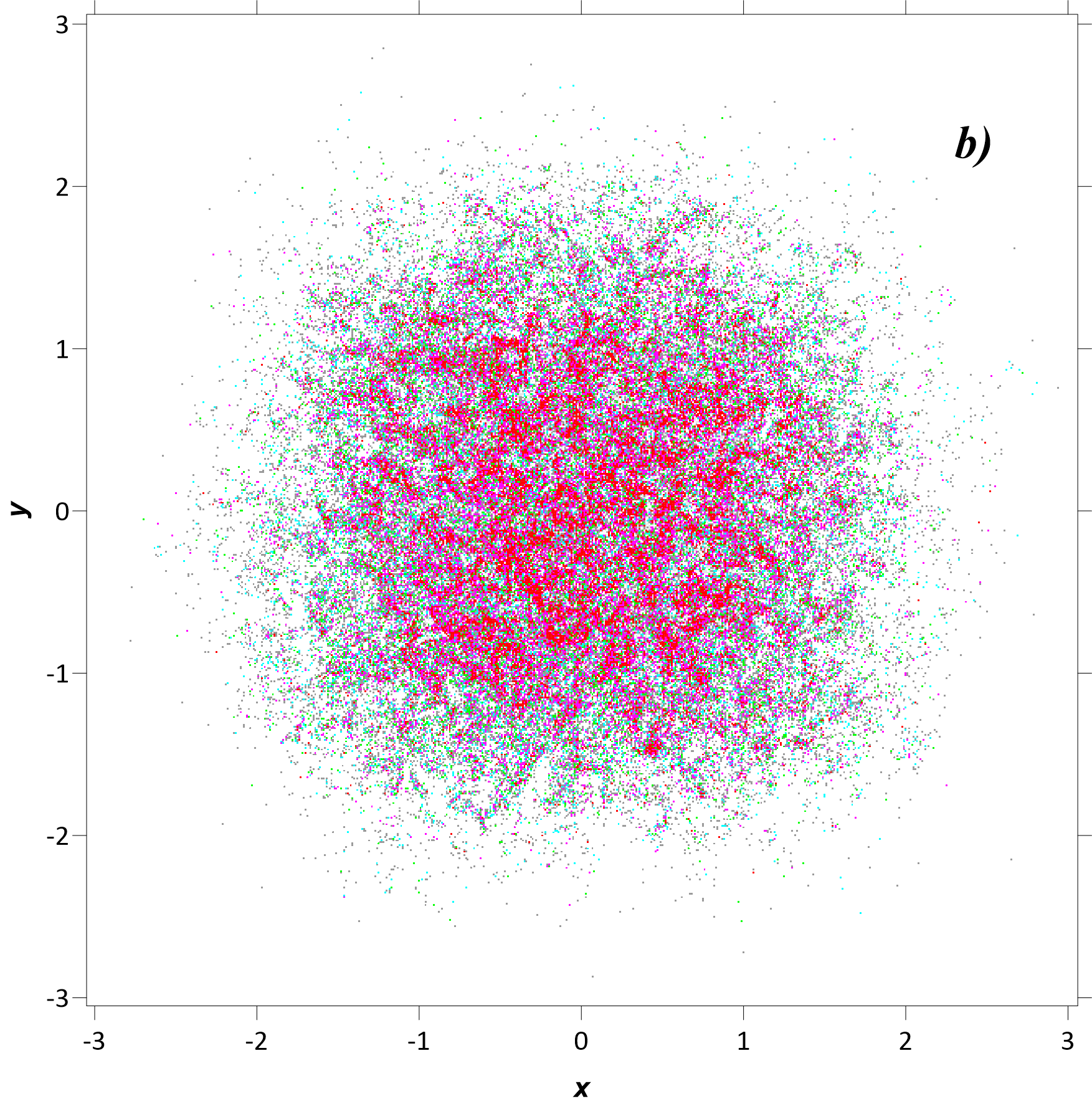}
\includegraphics[width=8.8 cm]{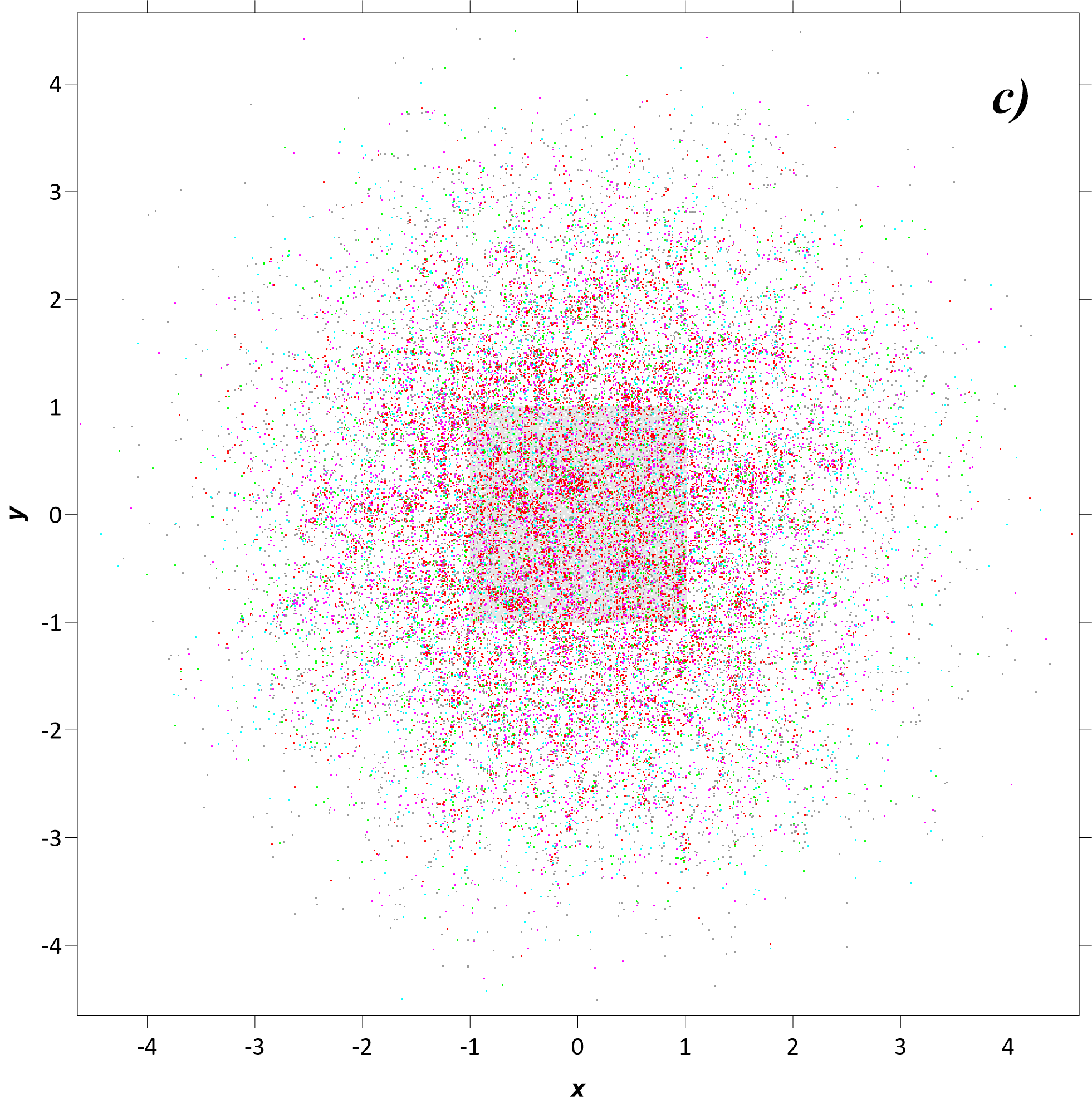}
\includegraphics[width=7.0 cm]{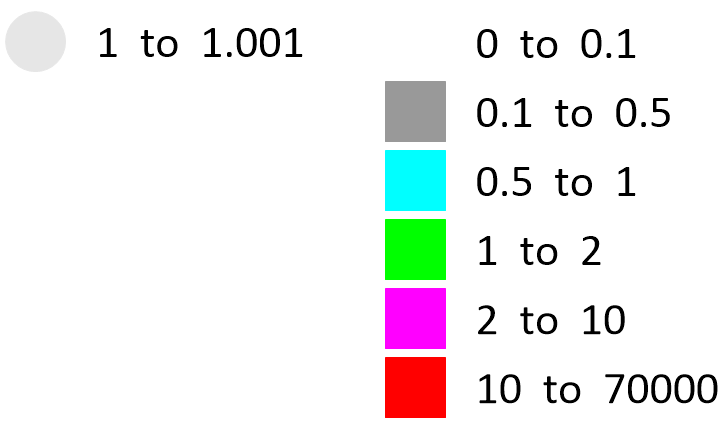}
\caption{
\label{fig10}
The same as Fig. \ref{fig8} but for $\gamma=0.1$.
}
\vskip20pt
\end{figure}

\begin{figure}
\centering
\includegraphics[width=6 cm]{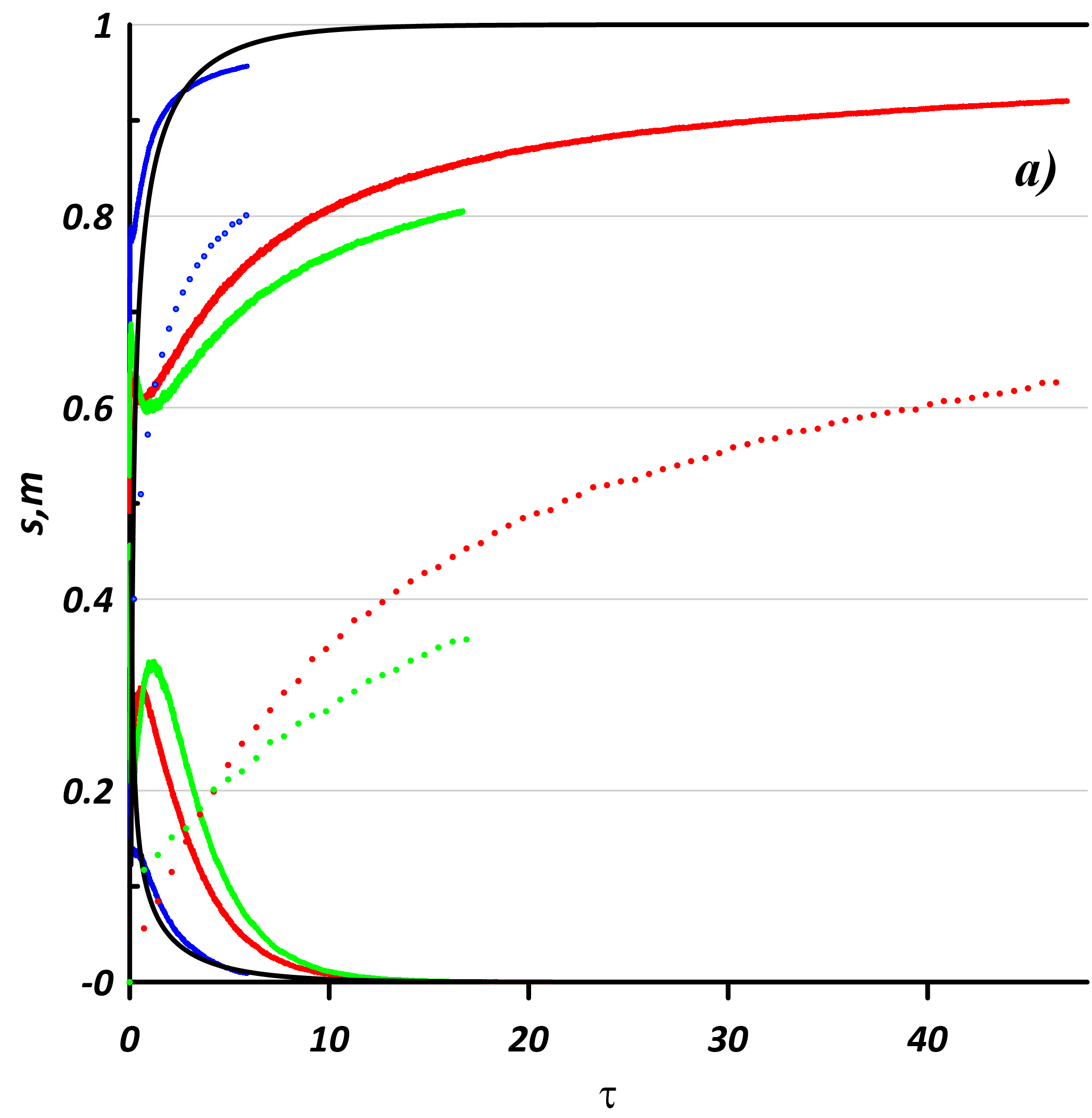}
\includegraphics[width=6 cm]{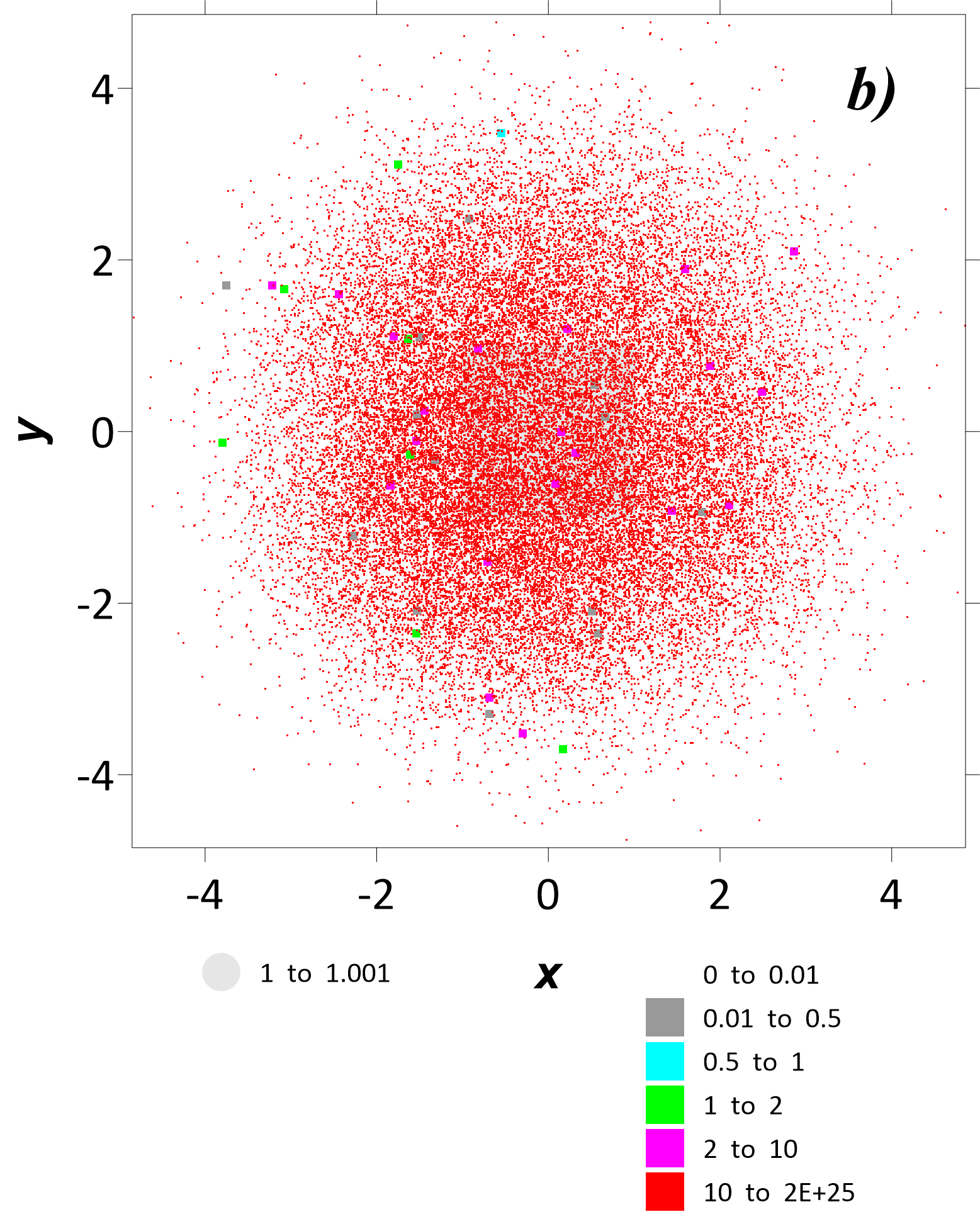}
\caption{
\label{fig11}
The same as Fig. \ref{fig7} but for $\gamma=0.1$.
}
\vskip20pt
\end{figure}

Let us now consider weakly divergent flow regimes with $\gamma=0.2$ (Fig. \ref{fig8}) and also $\gamma=0.1$ to illustrate the trends.
Because of the strong solenoidal component, the particles are faster scattered away from their initial
positions, hence we had to increase the model domain.
Although the asymptotic theory predicts no complete clustering, we have found that clustering still takes place, but at a significantly slower rate.
Figure \ref{fig9}a shows that initial clustering is faster due to the fragmentation process, but then it slows down.
The complete clustering limit is reached at about $t=25\tau$.
Stirring induced by the solenoidal flow component is significant, and in accord with this: $D^s \approx 16 D^p$.
Concurrently, the clustering process evolves over the longer solenoidal diffusion time scale $D^s$
(Fig. \ref{fig9}).
Clusters on the scale $l_{corr}$ do not emerge due to the shearing effect of the solenoidal component, which facilitates fragmentation of the clusters.
Complete clustering still occurs, even for $\gamma=0.1$ (Figs. \ref{fig10} and \ref{fig11}), but with 
progressively slower rates.
Dependence on $l_{corr}$ is similar to the $\gamma=0.2$ regime, as manifested by the clustering curves.

\begin{figure}
\centering
\includegraphics[width=6 cm]{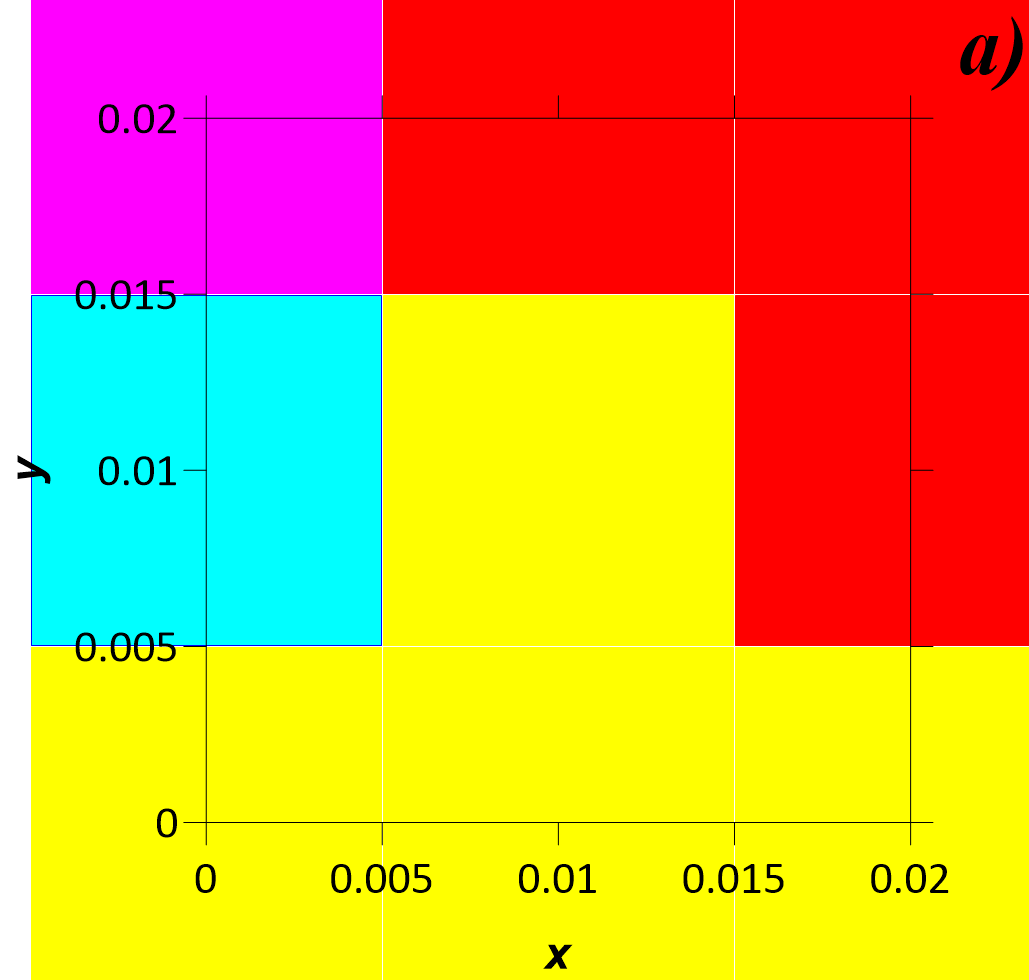}
\includegraphics[width=6 cm]{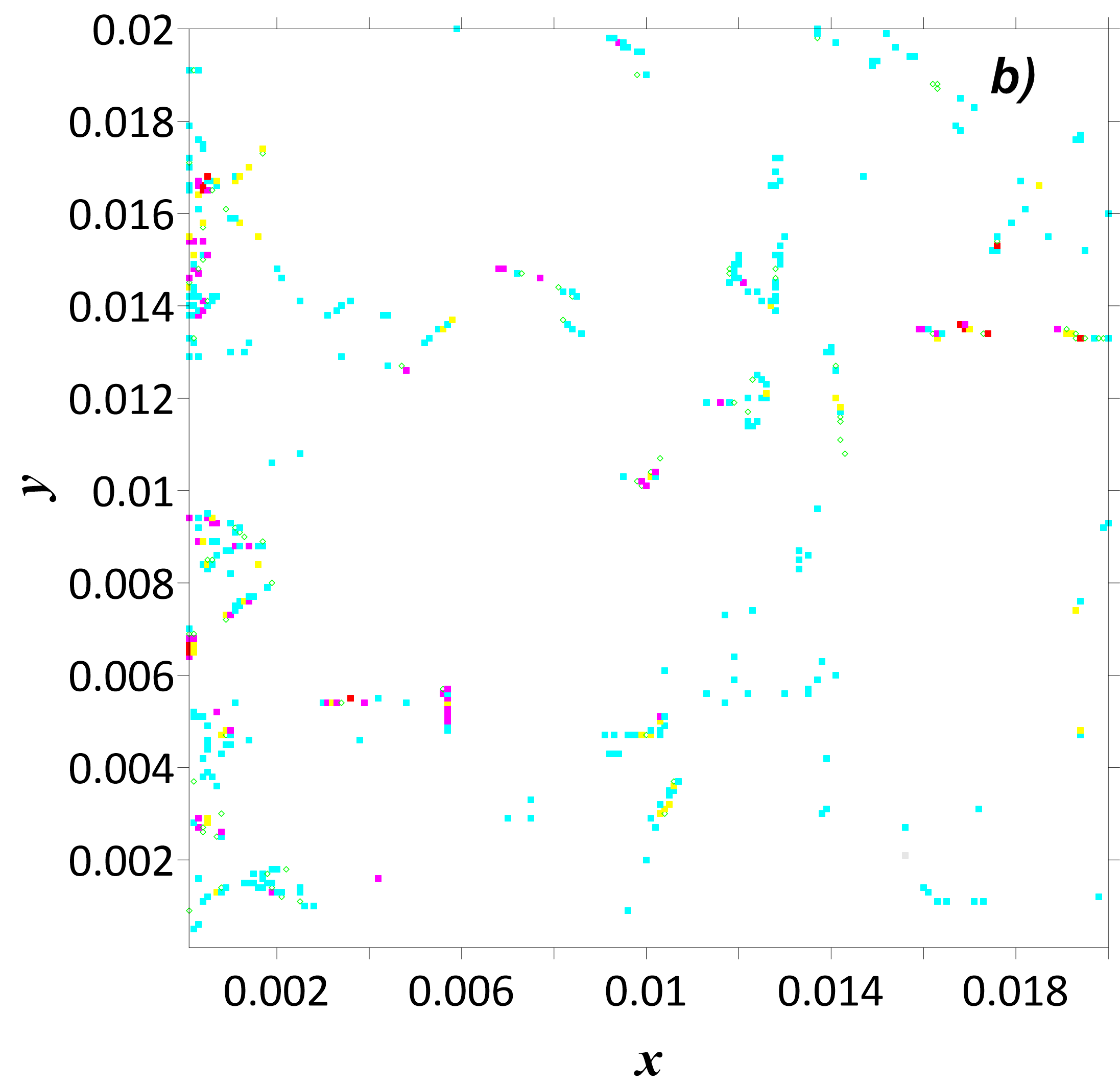}
\includegraphics[width=6 cm]{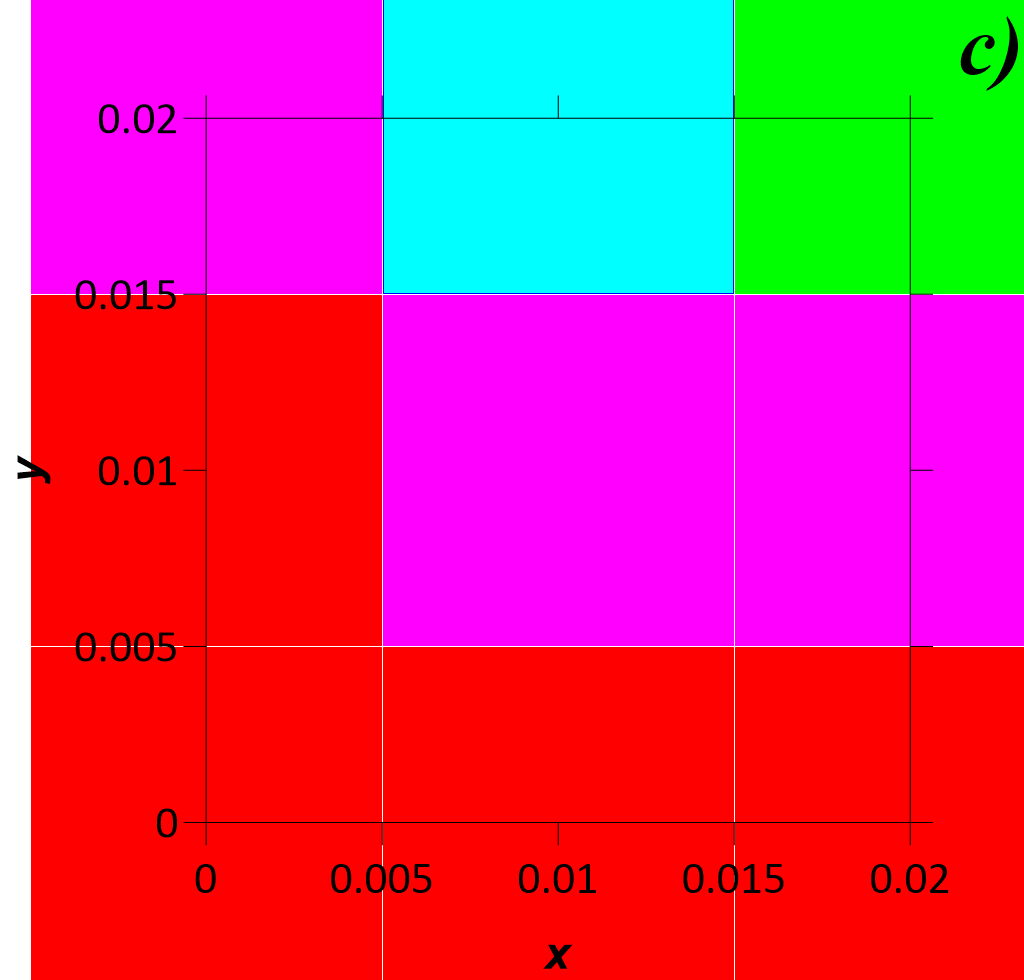}
\includegraphics[width=6 cm]{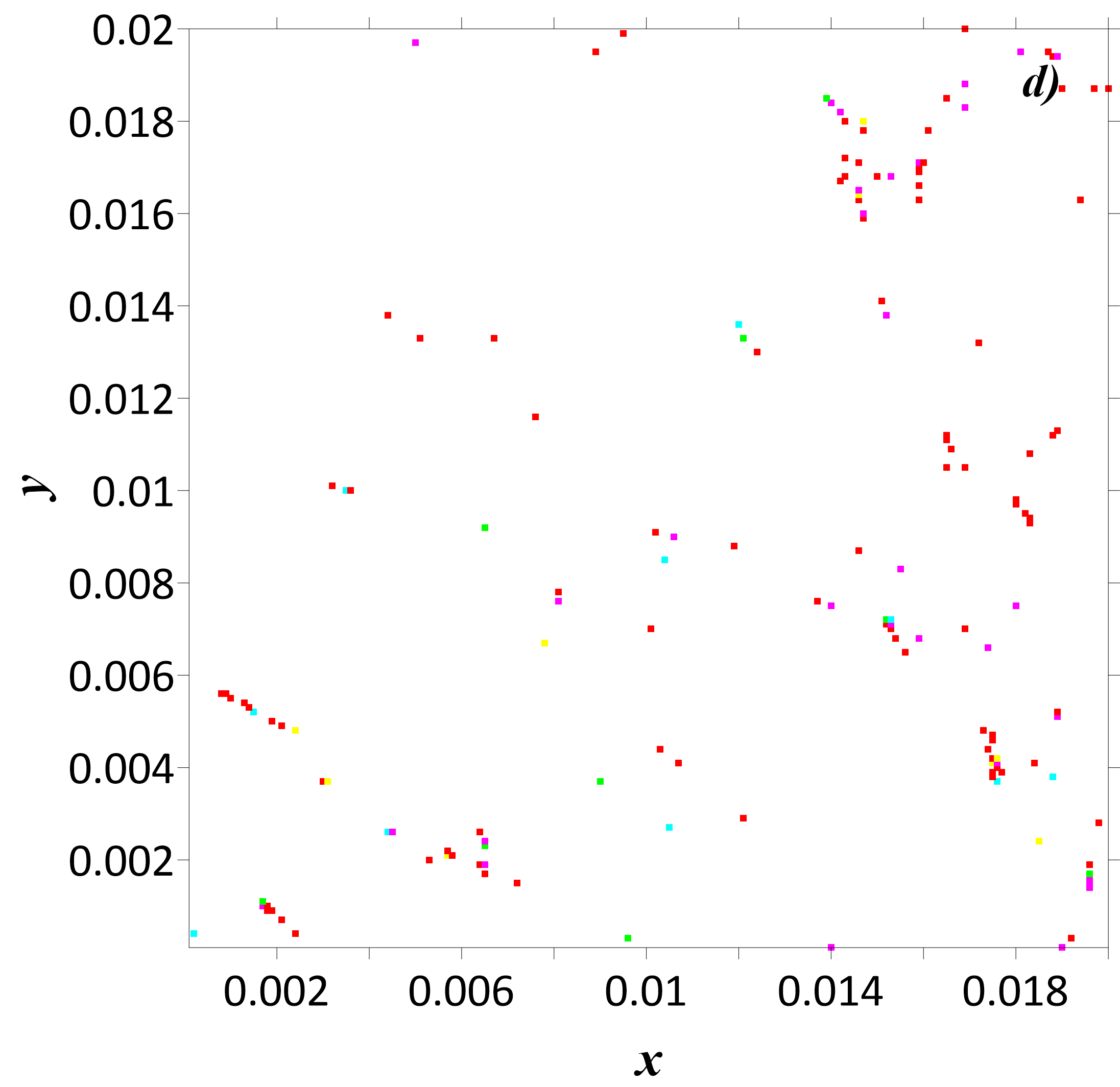}
\caption{
\label{fig101}
Effect of the coarsening used to obtain concentration.
Density distributions for $\gamma=0.2, \hskip5pt l = 0.04, \hskip5pt \sigma_U=0.33$.
Panels (a) and (c) show subdomain zoomed in from Fig. \ref{fig8}; panels (b) and (d) show the same but with
100 times finer resolution.
In (a) and (b) $t=0.422\tau$; in (c) and (d) $t=2.322\tau$.
}
\vskip20pt
\end{figure}

Let us now discuss effects of spatial averaging (i.e., coarse-graining) on the concentration.
If the averaging is too coarse, it may appear that all particles within the averaging region are completely clustered 
%***THESE ARE STRANGE WORDS, AS IT APPEARS COMPLETELY OPPOSITE***; 
while the corresponding clustering mass curves have not completely saturated yet (see Fig. \ref{fig9},
where the clusters are already red-colored).
To explain this, let us consider in detail the regime $\gamma=0.2$ and refine its spatial averaging by $100$ times in each direction (Fig. \ref{fig101}). {\er The finer structure of clusters, otherwise smeared out by the averaging, becomes apparent. Hence, within each coarse-grained cluster there are in fact many finer clusters that can be revealed by refining the averaging scale.}
The refined distributions show that many particles have not yet clustered, and this is true even for  relatively long times.
To summarize, some caution is required to interpret coarse-grained concentration and density fields, and in
this regard statistical topography diagnostics are useful.

%
%%%%%%%%%%%%%%%%%%%%%%%%%%%%%%%%%%%%%%%% CONCLUSIONS %%%%%%%%%%%%%%%%%%%%%%%%%%%%%%%%%%%%%%%%
%
\section{Conclusions}

This study is motivated by clustering of floating tracers and material objects on the ocean surface.
We consider both passive and floating (i.e., buoyant) tracers and show that their dynamics at the surface
are fundamentally different.
This is because passive-tracer concentration is materially conserved (i.e., remains constant on fluid particles),
whereas floating-tracer density (we use this term to distinguish it from the passive-tracer concentration)
changes due to the compressibility effect induced by the surface flow convergence.
Our main focus is on the density but some comparisons with the concentration are made for the sake of better understanding.

We represent tracers by ensembles of Lagrangian particles and consider their evolution in random kinematic
velocity fields.
The goal is to establish and interpret the clustering properties induced by 2D velocity fields consisting of divergent (potential) and non-divergent (solenoidal) flow components.
In the situation when velocity field is purely divergent (i.e., without a solenoidal component) and delta-correlated
in time, the existing asymptotic theory \citep{Klyatskin_2015}, which forms the basis of our work, predicts
complete clustering, that is, formation of spatial singularities that aggregate all available tracer.
The complete clustering develops exponentially in time and depends on the velocity correlation length scale.
If the divergent flow component is weaker than the solenoidal one, the asymptotic theory predicts no complete
clustering --- this regime has never been explored.
In this paper we close this gap by direct numerical modelling of the tracer fields and subsequent analysis
of the solutions by the means of statistical topography methodology.

The presented results demonstrate the opposite to the asymptotic prediction: the complete clustering is still
in place but its development becomes significantly slower in time.
The time scale in this case is controlled by the solenoidal flow component and its associated diffusion time
scale.
We also have found that in weakly divergent flow regimes, there is another type of transient tracer aggregation, referred to as fragmentation clustering, which coexists with the complete clustering and interferes with it.
This type of clustering is found to be bounded by finite values of concentration (typically $20$-$30$ times larger
than the initial one), and it is controlled by the solenoidal flow component that disperses particles and,
thus, inhibits the complete clustering.
In this regime clustering depends on the velocity correlation length scale.

The analysis and results of this paper should be of interest to ocean and atmosphere modelers, because realistic
ocean surface flows are weakly divergent due to ageostrophic motions.
Future research extensions considering effects of large-scale flows, explicit eddy diffusion, finite life time
of the tracer and inertial effects due to the buoyancy and finite size of floating objects.
Including finite time correlations in the kinematic flow model, as well as considering dynamically constrained
and progressively more realistic flows are also obvious avenues for new advances on the clustering problem.

%
%%%%%%%%%%%%%%%%%%%%%%%%%%%%%%%%%%%%%%%% ACKNOWLEDGMENTS %%%%%%%%%%%%%%%%%%%%%%%%%%%%%%%%%%%%%%%%
%
\begin{acknowledgments}
This study was supported by the POI FEB RAS Program "Mathematical simulation and analysis of dynamical processes
in the ocean" ($117030110034-7$) and by the Russian Foundation for Basic Research project $17-05-00035$.
EAR and PB were supported by the NERC grant $NE/R011567/1$ and by the Royal Society Exchange Grant
$IEC/R2/181033$.
KVK work on the statistical topography analysis was supported by the Russian Scientific Foundation project $9-17-0006$.
DVS was supported by the Russian Foundation for Basic Research project $19-55-10001-KO-a$.
VIK was supported by the Program of Fundamental Studies of the Russian Academy of Sciences "Advanced methods of mathematical modelling for studying nonlinear dynamical systems".
\end{acknowledgments}

\textbf{Appendix A}

Given the acting 3D velocity field $ {\bf u} \= (u,v,w)$, the continuity equation for the passive-tracer
concentration $ C_{3D}(x,y,z,t) $ is
\[
\pd{C_{3D}}{t} +\pd{(u \, C_{3D})}{x} +\pd{(v \, C_{3D})}{y} +\pd{(w \, C_{3D})}{z} =0 \, .
\]
Let us restrict our dynamical description to the 2D surface of the ocean and assume the rigid-lid approximation:
$ w(x,y,0,t) \= 0, $ with the surface velocity expressed as $ {\bf U} \= (U,V). $
In this case, the 2D governing equation for the surface concentration $ C(x,y,t) \= C_{3D}(x,y,0,t) $
can be written as
\[
\pd{C}{t} +{\bf U} \tochka \nabla C +C \, \nabla \tochka {\bf U} +C \, \pd{w}{z} \Big |_{z=0} =0 \, .
\]
Now, let us invoke the incompressible fluid density continuity equation $ \nabla \tochka {\bf u} \= 0, $
and express
\[
\pd{w}{z} \Big |_{z=0} =-\nabla \tochka {\bf U} \, .
\]
By taking this into account, the governing equation for evolution of the passive-tracer concentration on the ocean surface becomes
\[
\pd{C}{t} +{\bf U} \tochka \nabla C =0 \, .
\]

Now, let us derive a similar equation for the floating-tracer density $ \rho $ by writing the standard
continuity equation
\[
\pd{\rho}{t} +\pd{(u \, \rho)}{x} +\pd{(v \, \rho)}{y} +\pd{(w \, \rho)}{z} =0 \, ,
\]
and noting that the vertical density flux $ w \rho \= 0, $ because we assume that the additional buoyancy
force completely balances and cancels out any vertical flux of the tracer, and the tracer is always restricted to stay on the ocean surface.
By taking this into account, the governing equation for evolution of the floating-tracer density
$ \rho(x,y,t) $ on the ocean surface becomes
\[
\pd{\rho}{t} +{\bf U} \tochka \nabla \rho +\rho \, \nabla \tochka {\bf U} =0 \, .
\]
Note, that the last term, which represents density compression by converging velocity, does not have its
counterpart in the governing equation for $ C$.

\bibliography{General}

\end{document}